\documentclass[12pt]{iopart}
\usepackage{graphicx}
\expandafter\let\csname equation*\endcsname\relax
\expandafter\let\csname endequation*\endcsname\relax
\usepackage{amsmath}
\usepackage{empheq}
\usepackage{esvect}
\usepackage{mathabx}
\usepackage{cite}
\usepackage{lscape}
\usepackage{varwidth}
\usepackage{color,soul}

\begin{document}

\title[Fifth force stiffness]{MICROSCOPE's constraint on a short-range fifth force}

\author{Joel Berg\'e$^1$, Martin Pernot-Borr\`as$^1$$^2$, Jean-Philippe Uzan$^3$$^4$, Philippe Brax$^5$, Ratana Chhun$^1$, Gilles M\'etris$^6$, Manuel Rodrigues$^1$, Pierre Touboul$^1$}

\address{$^1$ DPHY, ONERA, Universit\'e Paris Saclay, F-92322 Ch\^atillon, France}
\address{$^2$ Sorbonne Universit\'e, CNRS, Institut d'Astrophysique de Paris, IAP, F-75014 Paris, France}
\address{$^3$ CNRS, Institut d'Astrophysique de Paris, IAP, F-75014 Paris, France}
\address{$^4$ Institut Lagrange de Paris, 98 bis, Bd Arago, 75014 Paris, France}
\address{$^5$ Institut de Physique Th\'eorique, Universit\'e Paris-Saclay, CEA, CNRS, F-91191 Gif-sur-Yvette Cedex, France}
\address{$^6$ Universit\'e C\^ote d{'}Azur, Observatoire de la C\^ote d'Azur, CNRS, IRD, G\'eoazur, 250 avenue Albert Einstein, F-06560 Valbonne, France}
\ead{joel.berge@onera.fr}
\vspace{10pt}
\begin{indented}
\item[]July 2020
\end{indented}

\begin{abstract}
The MICROSCOPE experiment was designed to test the weak equivalence principle in space, by comparing the low-frequency dynamics of cylindrical ``free-falling" test masses controlled by electrostatic forces. We use data taken during technical sessions aimed at estimating the electrostatic stiffness of MICROSCOPE's sensors to constrain a short-range Yukawa deviation from Newtonian gravity. We take advantage of the fact that in the limit of small displacements, the gravitational interaction (both Newtonian and Yukawa-like) between nested cylinders is linear, and thus simply characterised by a stiffness. By measuring the total stiffness of the forces acting on a test mass as it moves, and comparing it with the theoretical electrostatic stiffness (expected to dominate), it is {\it a priori} possible to infer constraints on the Yukawa potential parameters. However, we find that measurement uncertainties are dominated by the gold wires used to control the electric charge of the test masses, though their related stiffness is indeed smaller than the expected electrostatic stiffness. Moreover, we find a non-zero unaccounted for stiffness that depends on the instrument's electric configuration, hinting at the presence of patch-field effects. Added to significant uncertainties on the electrostatic model, they only allow for poor constraints on the Yukawa potential. This is not surprising, as MICROSCOPE was not designed for this measurement, but this analysis is the first step to new experimental searches for non-Newtonian gravity in space.
\end{abstract}

%

\submitto{\CQG}
%
%
%

\section{Introduction}

A hundred years after its invention, Einstein's theory of General Relativity (GR) still passes all experimental tests \cite{will14}, from early tests (the Mercury perihelion puzzle and the measurement of the gravitational deflection of stars' light passing near the Sun by Eddington) to current tests (gravitational lensing \cite{bartelmann01, hoekstra08}, gravitational redshift \cite{delva18, herrmann18}, gravitational waves direct detection \cite{abbott16}).
However, in order to withstand (not so recent) astrophysical and cosmological observations, GR must be supplemented by dark matter and dark energy. The former explains the flat rotation curve of galaxies and their dynamics in clusters \cite{zwicky33, rubin70}, while the latter explains the acceleration of the cosmic expansion \cite{riess98, perlmutter99}. Whether our theory of gravitation must be revised or the content of our Universe better understood is still an open discussion \cite{joyce15, joyce16}. In this article, we adopt the former possibility.

Theories beyond the standard model propose the existence of new fields and particles. For instance, string-inspired theories introduce a spin-0 dilaton-like particle (see e.g. Refs. \cite{damour94, damour02}), while scalar-tensor models modify GR's equations via the introduction of a new scalar field (see e.g. Refs. \cite{damour92, clifton12, joyce15}). 
Although a new very light scalar field should entail the appearance of a new long-range force incompatible with current Solar System tests, its existence can be made compatible with experimental constraints by virtue of a screening mechanism that makes the field's mass environment-dependent, thereby hiding it from local experimental tests \cite{damour94,vainshtein72,Damour:1992kf,khoury04a, khoury04b, babichev09,hinterbichler10, brax13,burrage18}. Those models can nevertheless have measurable effects, such as an apparent violation of the equivalence principle (see e.g. Refs. \cite{khoury04b, damour12}) or a variation of fundamental constants \cite{uzan03, uzan11}.

Looking for short-range deviations from Newtonian gravity is essential to test low-energy limits of high-energy alternative theories (such as string theory or extra dimensions) and is the goal of several experimental efforts (see Refs. \cite{fischbach99, adelberger03, adelberger09} for reviews and references therein, and Refs. \cite{tan20, lee20} for recent results). While most of them are highly optimised to look for specific minute signals, we aim, in this article, to search for a short-range deviation from Newtonian gravity as a byproduct of MICROSCOPE data.

The MICROSCOPE space experiment tested the weak equivalence principle (WEP) to a record accuracy \cite{touboul17, touboul19} via the comparison of the acceleration of two test masses freely falling while orbiting the Earth. If the WEP is violated, a signal is expected at a frequency defined as the sum of the satellite's orbital and spinning frequencies. Since MICROSCOPE orbits the Earth at a 700 km altitude, the experiment is then mostly sensitive to long-ranged (more than a few hundred kilometers) modifications of gravitation.  Its first results thus allowed us to set new limits on beyond-GR models involving long-range deviations from Newtonian gravity parametrised by a Yukawa potential, a light dilaton \cite{berge18} and a U-boson \cite{fayet18,fayet19}. Updates of those works are under way following the final MICROSCOPE results \cite{mic20, metris20}.

In this article, we use MICROSCOPE sessions dedicated to the in-flight characterisation of its instrument to look for short-range deviations of Newtonian gravity. Although in WEP-test configuration, the MICROSCOPE test masses are kept almost motionless by the electrostatic measurement apparatus and are (by design) barely affected by the satellite's and instrument's self-gravity, this is not the case in some technical sessions where they are set in a sinusoidal motion. In this situation, they are sensitive to other forces such as the instrument's electrostatic stiffness and the gravitational force from other parts of the instrument. Given the geometry of the MICROSCOPE instrument, we can expect to see deviations from Newtonian gravity ranging from the millimetre to the decimetre scales once all other environmental interactions are accounted for (because gravity tests are degenerate with the environment, as shown e.g. in Ref. \cite{berge18cqg}). Nevertheless, we must note that MICROSCOPE was not designed for this experiment, and we cannot expect to obtain competitive results. 
This paper is a prospective analysis of this new experimental concept so that its primary intent is to present this idea, and show that it should be substantially improved to test short-range gravity in space, hence allowing us to discuss its feasibility and limits.

The layout of this paper is as follows. We briefly introduce the Yukawa deviation from Newtonian gravity in Sect. \ref{sect_yukawa}.
Section \ref{sect_mic} gives a primer about MICROSCOPE's experimental concept; we present the equations of the dynamics of the test masses that are used in the remainder of the paper to measure an overall stiffness and infer constraints on a Yukawa interaction.
Section \ref{sect_forces} presents the specific experiment used in this work to measure the stiffness, and provides an exhaustive account of the forces at play, some of which are estimated from the data in Sect. \ref{sect_analysis}.
We present the data analysis procedure and the measurement of relevant parameters in Sect. \ref{sect_analysis}. We then provide new (albeit not competitive) constraints in Sect. \ref{sect_results}, before concluding in Sect. \ref{sect_conclusion}. Appendices give a pedagogical derivation of the electrostatic force at play along MICROSCOPE's cylinders' radial axes, and an analytical expression for the gravitational (both Newtonian and Yukawa) interaction between two cylinders.

\section{Yukawa gravity} \label{sect_yukawa}

We parametrise a deviation from Newtonian gravity with a Yukawa potential, which is simply added to the Newtonian potential. The total gravitational potential created by a point-mass of mass $M$ at distance $r$ is then
\begin{equation} \label{eq_yukawa_pm}
V(r) = -\frac{GM}{r} \left[1+\alpha \exp\left(-\frac{r}{\lambda}\right) \right],
\end{equation}
where $G$ is Newton's gravitational constant, $\alpha$ is the strength of the Yukawa deviation compared to Newtonian gravity and $\lambda$ is the range of the corresponding fifth force.

Despite its simplicity, the Yukawa parametrisation is useful as it describes the fifth force created by a massive scalar field in the Newtonian regime (see e.g. the Supplemental material of Ref.~\cite{berge18} and references therein).
The range $\lambda$ corresponds to the Compton wavelength of the scalar field, and $\alpha$ is linked to its scalar charge. Phenomenologically, this charge can depend on the composition of the interacting bodies in various ways, e.g. through combinations of their baryon and lepton numbers \cite{fischbach99}. In this paper, we consider composition-independent Yukawa interactions only (thereby, we assume a universal scalar charge), and we do not relate to any phenomenological subatomic model, but instead consider only $\alpha$ as the parameter to constrain.

Many experiments have already provided tight constraints on its range and strength, from sub-millimeter to Solar System scales (e.g. Refs.~\cite{adelberger03,fischbach99} and references therein, and Refs.~\cite{kapner07,masuda09,sushkov11,klimchitskaya14,berge18} for more recent works). In this article, we are concerned with ranges between $\lambda\approx 10^{-3}~{\rm m}$ and $\lambda\approx10^{-1}~{\rm m}$, corresponding to the scale of MICROSCOPE's instrument. The best constraints on the strength of a Yukawa potential for such ranges are $|\alpha|\leqslant 10^{-3}$ \cite{yang12, tan16, tan20}.

\section{MICROSCOPE experiment concept and test masses dynamics} \label{sect_mic}

MICROSCOPE was designed as a test of the universality of free-fall, relying on an easy recipe: drop two test bodies and compare their fall. However, instead of letting two test bodies freely orbit the Earth and monitoring their relative drift, MICROSCOPE uses electrostatic forces to maintain two test masses centered with respect to each other. This is done with a differential ultrasensitive electrostatic accelerometer, consisting of two coaxial and concentric cylinders made of different materials. The difference of electric potentials applied to keep the cylinders in equilibrium is a direct measure of the difference in their motion. 

This section provides a primer about the MICROSCOPE experimental concept. We start with a description of the capacitive detection and electrostatic control principle, driving us to a short presentation of the instrument. We then present a simplified equation of the dynamics of a test mass. Details can be found in Refs \cite{touboul20, liorzou20, robert20}.

\subsection{Electrostatic measurement principle}

The electrostatic control of the test masses relies on two nested control loops. 
The first one is inside the payload: each test mass is placed between pairs of electrodes and its motion with respect to its cage is monitored by capacitive sensors. It can be kept motionless by applying the electrostatic force required to compensate all other forces, such that the knowledge of the applied electrostatic potential provides a measurement of the acceleration which would affect the test mass with respect to the satellite in the absence of the control loop. Note that even if there is no net motion with respect to the satellite, it is common and convenient to call electrostatic acceleration the electrostatic force divided by the mass; this definition will be used all along the paper.
The second control loop is included in the satellite's Drag-Free and Attitude Control System (DFACS), which aims to counteract external disturbances via the action of cold gas thrusters. 
This system also ensures a very accurate control of the pointing and of the attitude of the satellite from the measurements of angular position delivered by the stellar sensors and of the angular acceleration delivered by the instrument itself.

Details about the former control loop can be found in Ref. \cite{liorzou20}. Nevertheless, to enlighten the discussion of the test masses' dynamics below, we provide some elements about the (central to the measurement) detection and action processes. The former allows for the measurement of the motion of the test mass, while the latter allows for its electrostatic control and the direct measurement of its acceleration.

A test mass surrounded by two opposite electrodes can be considered as a double capacitor with capacitances $C_1$ and $C_2$. The motion of the test mass induces a variation in the capacitances; the detector senses it and outputs a related voltage $V_{\rm det} = 2V_d (C_1 - C_2) / C_{\rm eq}$, where $V_d$ is the potential of the test mass and $C_{\rm eq}$ is the capacitance of the capacitor formed by the test mass and the electrodes when the test mass is at the centre of the cage.
The capacitances $C_i$ depend on the geometry of the sensor, and therefore their form differ along the longitudinal and radial axes of the instrument; nevertheless, it can be shown that at first order, $V_{\rm det}$ is proportional to the displacement $\delta$ of the test mass about the center of the cage along all axes \cite{liorzou20}.

The control loop digitises the detector output voltage $V_{\rm det}$ and computes the actuation voltage to apply to the electrodes in order to compensate for the displacement of the test mass, and recentre it in the cage. The (restoring) electrostatic force applied by an electrode $i$ is then
\begin{equation}
\vv{F}_{{\rm el}, i} = \frac{1}{2} (V_i - V_p)^2 \vv\nabla C,
\end{equation}
where $\vv\nabla C$ is the spatial gradient of the capacitance. The (polarisation) potential of the test-mass $V_p$ is maintained constant and the potential $V_i$ of the electrode is tuned by the servo-control loop.
This ``action" takes the general form
\begin{equation} \label{eq_action}
F_{{\rm el}, i} \approx -m \left\{G_{\rm act} V_e + \omega_p^2 \left[1 + \left(\frac{V_e}{V_p}\right)^2\right] \delta \right\},
\end{equation}
where the sensitivity factor $G_{\rm act}$ and the stiffness coefficient $\omega_p^2$ depend on the geometry of the sensor, $V_e$ is the voltage output from the control loop and applied to the electrodes, and $m$ is the mass of the test mass.

If $G_{\rm act}$ is known well enough, the acceleration of the test-mass can be measured through the voltage $V_e$ required to apply the restoring force. This measurement is perturbed by the electrostatic stiffness, which introduces a bias if the test-mass is not servo-controlled to the equilibrium point. Nevertheless, the asymmetry in the design of the electrostatic configuration and the displacement are sufficiently small to ignore it during nominal WEP test operations. Instead, in this paper, we use measurement sessions where the displacement $\delta$ is not negligible, allowing for the estimation of the electrostatic stiffness.

\subsection{Instrumental apparatus}

The core of MICROSCOPE's instrument consists of two differential accelerometers (or Sensor Units -- SU), the test masses of which are co-axial cylinders kept in equilibrium with electrostatic actuation \cite{liorzou20}. The test masses' materials were chosen carefully so as to maximize a potential violation of the WEP from a light dilaton \cite{damour94, damour10a, damour10b} and to optimise their industrial machining: the SUEP (Equivalence Principle test Sensor Unit) test masses are made of alloys of platinum-rhodium (PtRh10 -- 90\% Pt, 10\% Rh) and titanium-aluminium-vanadium (TA6V -- 90\% Ti, 6\% Al, 4\% V), while the SUREF (Reference Sensor Unit) test masses are made of the same PtRh10 alloy. For each sensor, we call ``IS1" the internal test mass and ``IS2" the external one. For instance, the internal mass of SUREF is named IS1-SUREF, and the external mass of SUEP is called IS2-SUEP.

As shown above, the test masses of each SU are controlled electrostatically with no mechanical contact but a thin 7 $\mu$m-diameter gold wire used to fix the masses' electrical potential to the electronics reference voltage. Two Front-End Electronics Unit (FEEU) boxes (one per SU) include the capacitive sensing of masses, the reference voltage sources and the analog electronics to generate the electrical voltages applied to the electrodes; an Interface Control Unit (ICU) includes the digital electronics associated with the servo-loop digital control laws, as well as the interfaces to the satellite's data bus. 

\begin{figure}
\begin{center}
\includegraphics[width=.9\textwidth]{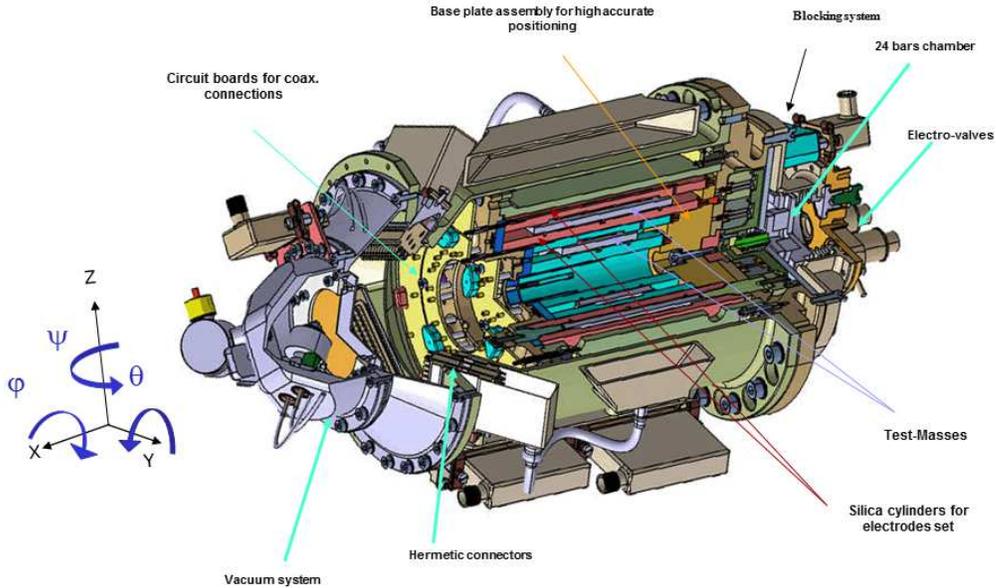}
\caption{Cutaway view of a MICROSCOPE sensor, with its two test masses, their surrounding electrodes-bearing cylinders, cylindrical invar shield, base plate, upper clamp and vacuum system. The volume of a sensor is $348 \times 180 \times 180~{\rm mm}^3$; the test masses have length ranging from 43~mm to 80~mm and outer radii ranging from 39~mm to 69~mm, and are separated from the electrode-bearing cylinders by 600~$\mu$m gaps \cite{liorzou20}. The reference system is shown on the left of the figure. Figure from Ref. \cite{touboul19, liorzou20}.}
\label{fig_sensor}
\end{center}
\end{figure}

Fig. \ref{fig_sensor} shows a cutaway view of one SU, with its two test masses, their surrounding electrodes-bearing cylinders, cylindrical invar shield, base plate, upper clamp and vacuum system.

\subsection{MICROSCOPE's test mass measured acceleration}

The dynamics of MICROSCOPE's test masses, as required for testing the WEP, is discussed at length in Ref. \cite{touboul20}.
In particular, Ref. \cite{touboul20} focusses on the differential motion of two test masses and discriminates between inertial and gravitational masses. In this section, we summarise the equations pertaining to the dynamics of a given test mass, and since this article is not concerned with the WEP, we identify the inertial and gravitational masses (thence some simplification with respect to Ref. \cite{touboul20}).

Up to electrostatic parasitic forces (see below), the electrostatic force (\ref{eq_action}) corresponds to a ``control" acceleration responding to the contribution of the various contributors to the dynamics of the test mass,
\begin{equation} \label{eq_measurement}
\vec\Gamma_{\rm cont} = \frac{\vec{F}_{\rm el}}{m} = \vv{\Delta\Gamma}_\Earth + \vec\Gamma_{\rm kin} - \frac{\vec{F}_{\rm loc}}{m} - \frac{\vec{F}_{\rm pa}}{m} + \frac{\vec{F}_{\rm ext}}{M}+ \frac{\vec{F}_{\rm th}}{M}
\end{equation}
where $m$ and $M$ are the masses of the test mass and of the satellite, $\vec{F}_{\rm ext}$ are non-gravitational forces affecting the satellite (atmospheric drag, Solar radiation pressure), $\vec{F}_{\rm th}$ are forces applied by the thrusters (to compensate for external forces) and $\vec{F}_{\rm loc}$ and $\vec{F}_{\rm pa}$ are local forces (inside the sensor) that we can consider individually (e.g. electrostatic stiffness, gold wire stiffness, self-gravity) or as collective contributions (e.g. electrostatic parasitic forces), respectively. We denote as $\vv{\Delta\Gamma}_\Earth$ the difference between the Earth gravitational acceleration at the center of the satellite and that at the center of the test mass.
We assume that the test masses are homogeneous. Moreover, since we are concerned with short-range Yukawa deviations only,  we assume that the Yukawa contribution to the Earth's gravity acceleration acting on the test-masses is negligible. 
Finally, the second term of the r.h.s. of Eq.~(\ref{eq_measurement}) contains the contribution from the satellite's inertia and from the motion of the test-mass,
\begin{equation} \label{eq_kinacc}
\vv\Gamma_{\rm kin} = [{\rm In}] \vv{P} + 2[\Omega] \dot{\vv{P}} + \ddot{\vv{P}},
\end{equation}
where $\vv{P}$ is the position of the test mass with respect to the center of the satellite, $[{\rm In}] \equiv [\dot\Omega] + [\Omega] [\Omega]$ is the gradient of inertia matrix of the satellite and $[\Omega]$ its angular velocity.

Noting that since the applied electrostatic force of Eq. (\ref{eq_measurement}) is the sum of the measured ``action" force (\ref{eq_action}) summed over all electrodes and parasitic electrostatic forces,
\begin{equation}
\vec{F}_{\rm el} = \vec{F}_{\rm el, meas} + \vec{F}_{\rm elec, par},
\end{equation}
we show in \ref{app_dynamics} that the measured acceleration of a test mass, expressed in the instrument frame, is
\begin{equation} \label{eq_main0}
\vv\Gamma_{\rm meas | instr} = \vv{B}_{0} + \vv{\Delta\Gamma}_{\Earth {\rm | sat}} + \vv\Gamma_{\rm kin | sat} - \frac{\vec{F}_{\rm loc | instr}}{m} + \vv{n},
\end{equation}
where $\vv{n}$ is the measurement noise and $\vv{B}_{0}$ is an overall bias defined from the local parasitic forces and measurement bias. 

\section{Stiffness: experimental measurement and contributors}  \label{sect_forces}

A short-range Yukawa fifth force may hide in the local component $\vec{F}_{\rm loc}$ of the force. As described in the remainder of the paper, our analysis is based on Eq. (\ref{eq_main0}) and consists in:
\begin{itemize}
\item measuring the overall stiffness using dedicated sessions;
\item estimating (when possible) or modelling all possible contributors to $\vec{F}_{\rm loc}$, then subtracting them (but a Yukawa interaction) from the measured overall stiffness;
\item extracting constraints on a Yukawa interaction from the residuals.
\end{itemize}

In this section, we first describe the experimental approach to measure the overall stiffness (subsection \ref{ssect_measurement}), before listing (subsection \ref{ssect_budget}) and discussing one by one the contributions to $\vec{F}_{\rm loc}$, including the Yukawa interaction, in the remaining subsections.
The second item of our programme is completed in this section (contributions that cannot be estimated and are thus modelled --electrostatic stiffness, thermal effects and Newtonian gravitational interaction) and in Sect. \ref{sect_analysis}, where contributions that can be extracted from the data (gold wire and Yukawa interaction) are estimated.
The last item is the subject of Sect. \ref{sect_results}.

\subsection{Electrostatic stiffness measurement sessions} \label{ssect_measurement}

The stiffness is the derivative of force with respect to the position.
Measurement sessions were dedicated to measure MICROSCOPE's instrument stiffness \cite{chhuncqg5}, the stiffness being expectedly dominated by an electrostatic stiffness (see Sect. \ref{ssect_electro} and Ref. \cite{liorzou20}). 
The principle of the measurement is to impart a $f=1/300$~Hz sinusoidal excitation of amplitude $x_0=5~\mu$m to the test mass through the electronics control loop.
The position of the test mass is thus forced to be
\begin{equation}
x(t) = x_0 \sin(\omega t + \psi),
\end{equation}
where $x(t)$ is any axis ($x,y,z$) of the instrument (along which we aim to estimate the electrostatic stiffness), $\omega = 2\pi f$ and $\psi$ a given phase.
The acceleration (\ref{eq_main0}) measured as the test mass motion is forced in position is 
\begin{equation} 
\vv\Gamma_{\rm meas} = \vv\Gamma_{\rm exc} + \vv{B}_{0} + \vv{\Delta\Gamma}_\Earth + \vv\Gamma_{\rm kin} - \frac{\vec{F}_{\rm loc}}{m} + \vv{n},
\end{equation}
where we dropped the subscripts ``$|$instr" and ``$|$sat" for simplicity, and where $\vv\Gamma_{\rm exc}~=~x_0~ \omega^2 ~\sin(\omega t + \psi)$ is the excitation acceleration imparted to the test mass from the electronics control loop.

\begin{figure}
\begin{center}
\includegraphics[width=.45\textwidth]{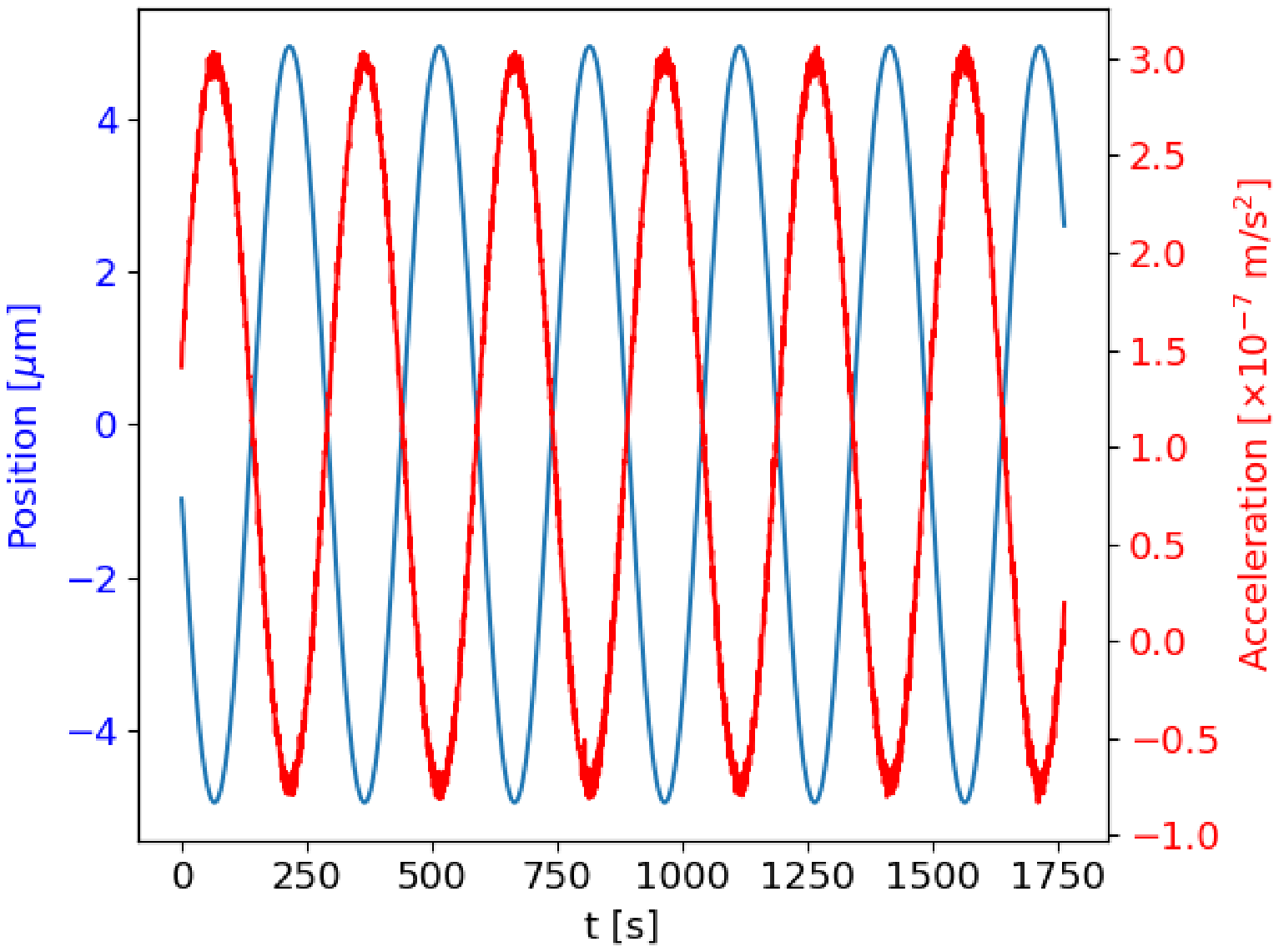}
\includegraphics[width=.45\textwidth]{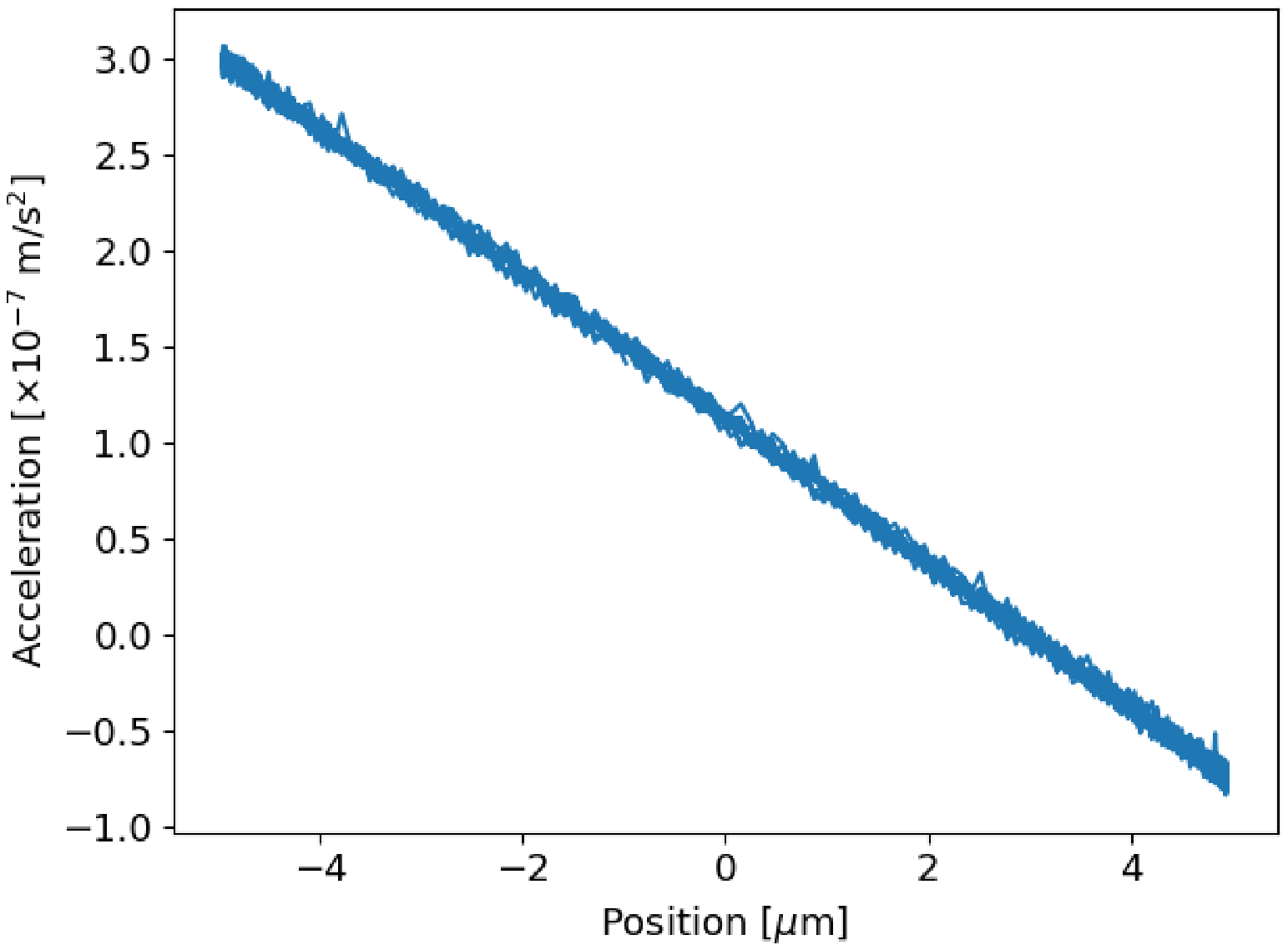}
\caption{Experimental process to measure a MICROSCOPE's sensor's stiffness (here, the stiffness of SUREF's internal mass is estimated along its $Y$ axis): the test mass is excited in position with a known amplitude of 5 $\mu$m, and we measure its response in acceleration (left panel); the acceleration is mostly due to the electrostatic stiffness. Right panel: test mass' acceleration as a function of its position.}
\label{fig_meas}
\end{center}
\end{figure}

Two measurements lasting 1750 s were performed for each axis of each test mass (one measurement per available electrical configuration --subsection \ref{ssect_electro} and \ref{app_hrmfrm}).
Figure \ref{fig_meas} shows the process to measure the stiffness of SUREF's internal mass along its $Y$ axis: the acceleration measured by the sensor (red) is compared to its input position (blue).

In the remainder of this paper, we subtract the Earth gravity modeled as described in Refs. \cite{touboul19, berge_cqg7}, as well as the kinematic acceleration inferred from satellite's attitude measurements, from the measured acceleration.
We thus deal with the acceleration
\begin{align} \label{eq_main1}
\vv\Gamma & \equiv \vv\Gamma_{\rm meas} - \vv{\Delta\Gamma}_\Earth - \vv\Gamma_{\rm kin} \\
\label{eq_main2}
&= \vv\Gamma_{\rm exc} + \vv{B}_{0} - \frac{\vec{F}_{\rm loc}}{m} + \vv{n}.
\end{align}

This acceleration is dominated by the local one $\vec{F}_{\rm loc}/m$, while the excitation acceleration $\vv\Gamma_{\rm exc}$ is negligible.
In Ref. \cite{chhuncqg5}, it was assumed that only the electrostatic stiffness $k_\epsilon$ played a significant role, such that $\vec{F}_{\rm loc} = -k_\epsilon \vec{x}$, where $\vec{x}$ is the displacement of the test mass with respect to its equilibrium position, and Eq. (\ref{eq_main2}) became (ignoring the quadratic factor)
\begin{equation} 
\vv\Gamma = \vv\Gamma_{\rm exc} + \vv{B}_0 + \frac{k_\epsilon}{m} \vec{x} + \vv{n}.
\end{equation}

Under these assumptions, the electrostatic stiffness is simply the slope of the $\vv\Gamma - \vec{x}$ relation (up to the factor $m$), as shown in the right panel of Fig. \ref{fig_meas}. Chhun {\em et al.} \cite{chhuncqg5} used this simple technique to estimate the electrostatic stiffness on the three axes of each MICROSCOPE's test mass. They found significant disagreements with expectations from the theoretical model summarised in Sect. \ref{ssect_electro}, which they explained by model inaccuracies and contribution from the gold wire aimed to control the charge of the test masses (Sect. \ref{ssect_gw}). We discuss their results in \ref{ssect_chhun}.

\subsection{Local forces budget} \label{ssect_budget}

In what follows, we go beyond the simple assumptions of Ref. \cite{chhuncqg5} and explore how the same measurement could shed light on short-range non-Newtonian gravity.

To that goal, we need to take into account all the different local forces applied to the test mass as it moves inside its electrode cage during sessions dedicated to the measurement of the stiffness. 
Our central equation is Eq. (\ref{eq_main2}), in which the local force is the sum of the following contributors discussed in this section:  electrostatic force $\vv{F}_{\rm el}$ (Sect. \ref{ssect_electro}); force due to the gold wire $\vv{F}_{\rm w}$ (Sect. \ref{ssect_gw}); radiation pressure $\vv{F}_{\rm p}$ (Sect. \ref{ssect_radp}); radiometric effect $\vv{F}_{\rm r}$ (Sect. \ref{ssect_radio}); Newtonian $\vv{F}_{\rm N}$ and  non-Newtonian (Yukawa) $\vv{F}_{\rm Y}$ gravity (Sect. \ref{ssect_grav}), such that $\vv{F}_{\rm loc} = \vv{F}_{\rm el} + \vv{F}_{\rm w} + \vv{F}_{\rm p} + \vv{F}_{\rm r} + \vv{F}_{\rm N} + \vv{F}_{\rm Y}$.

\subsection{Electrostatic force} \label{ssect_electro}

The electrostatic force used to control the test mass is discussed at length in Ref. \cite{liorzou20}. Here, we shall only state that it consists of a bias $\vv{b}_\epsilon$ and a stiffness $k_{\epsilon}$,
\begin{equation}
\vv{F}_{\rm el} = \vv{b}_\epsilon - k_{\epsilon} \vv{x}.
\end{equation}

Those factors depend on the geometry of the test mass and of the electrodes, and on the electric configuration (voltages applied to the different parts of the sensor). In particular, the electrostatic stiffness along the $X$-axis is expected to be zero for all sensors.
For completeness, and since this paper particularly focuses on the stiffness, we provide below the electrostatic force imparted by the full set of electrodes on the radial axes when the test mass moves along the $Y$-axis \cite{liorzou20}:
\begin{multline} \label{eq_forcey}
F_{\rm el}(y) \approx -\frac{4\epsilon_0 S_y}{e_i^2} \frac{\sin (\alpha_y/2)}{\alpha_y/2} (V'_{p_y}-V_p)v_y \\
+ \frac{2 \epsilon_0 S}{e_i^3} \left(1 + \frac{\sin \alpha_y}{\alpha_y} \right) [(V'_{p_y}-V_p)^2 +V_d^2]y \\
+ \frac{2 \epsilon_0 S}{e_i^3} \left(1 - \frac{\sin \alpha_z}{\alpha_z} \right) [(V'_{p_z}-V_p)^2 +V_d^2]y \\
+\frac {2\pi \epsilon_0 L_x R_x} {e_e^3} [(V'_{p_x}-V_p)^2 + V_d^2]y \\
+\frac {\pi \epsilon_0 R_{\phi} L_{\phi}} {e_e^3} [(V'_{p_{\phi}}-V_p)^2 + V_d^2]y,
\end{multline}
where $\epsilon_0$ is the vacuum permittivity, $e_i$ ($e_e$) is the gap between the inner (outer) electrode cylinder and the test-mass, and where we assumed that all control voltages  listed in Ref. \cite{liorzou20} are small compared to the $V_p$ and $V_d$ voltages. Those two voltages describe the electric configuration. Two configurations are available: high-resolution mode (HRM) and full-range mode (FRM). They are detailed in Ref. \cite{liorzou20} and summarised in \ref{app_hrmfrm}. 

The first term of the r.h.s. of Eq. (\ref{eq_forcey}) defines the gain of the detector (the force being proportional to the control voltage $v_y$); the other terms define the stiffness created by the $Y$, $Z$, $X$ and $\phi$ electrodes. In this equation, $S$ is the surface of the $Y$ and $Z$ electrodes, $R_x$ and $R_\phi$ are the inner radius of the $X$ and $\phi$ electrodes, and $L_x$ and $L_\phi$ are their length. The angles $\alpha_y$ and $\alpha_z$ are defined as the angle between the displacement of the test mass and the $Y$ and $Z$ axes, respectively. 
\ref{app_estiffness} proves the form of the stiffness created by the $Y$ electrodes (second term of the r.h.s. of the equation).

We assessed the accuracy of the stiffness terms of the model \eqref{eq_forcey} with finite elements simulations. We found it to be biased high: finite elements models provide an electrostatic stiffness 7\% to 10\% lower than the model \eqref{eq_forcey}. Nevertheless, in the remainder of this paper, instead of relying on finite elements simulations, we use Eq.~(\ref{eq_forcey}) corrected by a 8.5\% bias. This allows us to easily propagate metrology and voltage uncertainties in the electrostatic stiffness model, without the need to run a time-consuming simulation for each allowed set of parameters. We then add an extra 3\% statistical error to those uncertainties to reflect the uncertainty on the bias of the model.
The 7th column of Table \ref{tab_results} (denoted $k_{\epsilon, {\rm th}}$) lists the electrostatic stiffness expected for each test mass of MICROSCOPE.

\subsection{Gold wire} \label{ssect_gw}

The electric charge on test masses is controlled via a gold wire linking them to the spacecraft body. The wire can be modelled as a spring acting on the test mass with the force
\begin{equation}
\vv{F_w} = -k_w [1+ {\rm i}\phi(f)] \vv{x} - \lambda_w \dot{\vv{x}},
\end{equation}
where $\lambda_w$ describes the viscous damping of the wire, $k_w$ is the wire stiffness and $\phi(f)$ describes the internal damping; note that $\phi$ can depend on the frequency $f$. The wire quality factor $Q=1/\phi$.

For a sinusoidal motion of the test mass (along the $j$th axis) $x_j(t) = x_{j0} \sin(\omega t + \psi)$, the force exerted by the gold wire is the sum of an out-of-phase sinusoidal signal \cite{saulson90} and a (velocity-proportional) quadrature signal
\begin{equation}
F_{w,j}(t) = -k_{w,j} x_{j0} \sin(\omega t + \psi - \phi) + \lambda_w x_{j0} \omega \cos(\omega t + \psi).
\end{equation}

Thermal dissipation in the wire is the origin of the $f^{-1/2}$ low-frequency noise that limits MICROSCOPE's test of the WEP \cite{touboul17}. 
Using the dissipation-fluctuation theorem, it can be shown that this acceleration noise is \cite{saulson90, willemenot00}
\begin{equation} \label{eq_goldwirePSD}
\Gamma_{n,w}(f) = \frac{1}{m} \sqrt{\frac{4k_BT}{2\pi} \frac{k_w}{Q(f)}}f^{-1/2} \,\, {\rm m s}^{-2} / \sqrt{\rm Hz},
\end{equation}
where $m$ is the mass of the test mass, $T$ is the temperature and $k_B$ is the Boltzmann constant.
This allows for an estimation of the $k_w/Q$ ratio from the spectral density of long measurement sessions (see Sect. \ref{ssect_kQ}).

\subsection{Radiation pressure} \label{ssect_radp}

The electrode-bearing cylinders, being at temperature $T$, emit thermal radiation through photons that eventually hit the test mass and transfer their momentum to it, thus creating a pressure. A gradient of temperature and a difference of temperature $\Delta T$ between the electrodes surrounding the test mass will therefore cause a force directed from the hottest to the coldest regions \cite{nofrarias07, carbone07}:
\begin{equation}
\vv{F}_{\rm p} = \frac{16}{3c} S \sigma \Delta T T^3 \vv{e},
\end{equation}
where $T$ is the average temperature, $c$ the speed of light, $\sigma$ the Stefan-Boltzmann constant, $S$ the surface of the test mass, and $\vv{e}$ is the vector directed from the hottest to the coldest region.

The temperature and its gradient did not evolve in time during the measurement sessions used in this paper (six temperature probes are positioned on each sensor in such a way that we can monitor the temperature and have a glimpse at its gradient \cite{hardycqg6}; in the worst case, we could note a 0.003K evolution of the temperature during the measurement --while its mean is about 280K--, with all probes affected by the same evolution, entailing an unmeasurably small variation of the temperature gradients). Therefore, as far as we are concerned, we can consider the radiation pressure-induced force as a simple bias. Given the measured temperatures, an order of magnitude estimation allows us to expect the corresponding acceleration to be at a level of $a_p \lesssim 10^{-10}~{\rm ms}^{-2}$.

\subsection{Radiometer effect} \label{ssect_radio}

Taking its name from Crookes' radiometer, originally thought to prove the photon pressure, the radiometer effect is actually a residual gas effect affecting test masses in rarefied atmospheres whose mean free path exceed the size of the container. In this case, equilibrium conditions do not happen when pressure is uniform, but when the ratios of pressure to square root of temperature equal one another \cite{nofrarias07, carbone07}.

This entails a force on the test mass proportional to temperature gradient about its faces $\Delta T$,
\begin{equation}
\vv{F}_{\rm r} = \frac{1}{2} PS \frac{\Delta T}{T} \vv{e},
\end{equation}
where $P$ is the pressure in the container, $S$ the surface of the test mass orthogonal to the temperature gradient, $T$ the average temperature in the container and, as before, $\vv{e}$ is the vector directed from the hottest to the coldest regions.

Even when stationary, a non-linear temperature profile can cause a position-dependent radiometric effect and potentially a stiffness. However, the sparse temperature measurements in MICROSCOPE sensors do not allow us to go beyond the linear temperature profile hypothesis, thereby limiting our modelling of the radiometric effect to a constant acceleration. 
Orders of magnitude estimates provide a level of acceleration of the same order as the radiation pressure, $a_r \lesssim 10^{-10}~{\rm ms}^{-2}$.

\subsection{Other non-gravitational effects}


\subsubsection{Residual gas drag} The test mass moves in an imperfect vacuum, so that drag may be expected. Orders of magnitude estimates provide a related acceleration $\approx 10^{-23}~{\rm ms}^{-2}$, well below our capacity to detect it \cite{liorzou20}.

\subsubsection{Outgassing} Gas molecules are released from the materials of the instrument's parts (in particular the electrode-bearing cylinders) and can impact the test mass and modify the pressure inside the instrument \cite{nofrarias07}. However, the vacuum system was designed, and the materials chosen, such that outgassing can be safely ignored \cite{liorzou20}.

\subsubsection{Lorentz force} Test masses have a non-zero magnetic moment, and can therefore be affected by Lorentz forces, either from the Earth magnetic field or local magnetic fields. The former applies a periodic signal at the orbital frequency and therefore does not affect the stiffness measurements (besides the fact that the Earth magnetic field is largely suppressed by MICROSCOPE's instrument magnetic shield). Local magnetic fields are more difficult to assess. However, noting that their effect on the test of the WEP is subdominant \cite{hardycqg6}, we ignore them in this paper.

\subsubsection{Contact-potential differences and patch fields} Inhomogeneous distributions of surface potentials create a force between charged surfaces. MICROSCOPE's instrument can be affected by such patch effects, which act as an additional stiffness dependent on the test masses' voltages, thus on the electric configuration \cite{speake96}. It goes beyond the scope of this article to develop a model of patch effects in MICROSCOPE, and we will not try to quantify them. Note that they may affect MICROSCOPE only in the stiffness measurement sessions used in this paper, where test masses are set in motion; in MICROSCOPE's test of the WEP, test masses are kept motionless, and thus less sensitive to patch effects (which act as a bias).

\subsubsection{Misalignments and geometrical defects} Very small misalignments between MICROSCOPE's cylinders can be estimated \cite{hardycqg6}. As they break the cylindrical symmetry, they can introduce additional terms in the electrostatic stiffness \cite{hudson07}. However, as we show below, the error budget in stiffness measurement sessions is largely dominated by the gold wire, so that we can safely ignore them for the purpose of this paper (thereby justifying our $[\theta_j] = {\rm Id}$ assumption in \ref{app_dynamics}).

\subsection{Local gravity} \label{ssect_grav}

The local gravity force applied to a MICROSCOPE test mass is the sum of the forces between that test mass and the parts making the corresponding sensor (Fig. \ref{fig_sensor}): 
\begin{itemize}
\item seven co-axial cylinders: two silica electrode-bearing cylinders surrounding the test mass, the second test mass and two other silica electrode-bearing cylinders surrounding it and two cylindrical invar shields, 
\item and four plain cylinders: a silica base plate, an invar base plate, an invar upper clamp, and a vacuum system.
\end{itemize}
The characteristics of those elements can be found in Ref. \cite{liorzou20}. As we show below, the gravity force is dominated by the closest elements, so that we can safely neglect the contribution from the other sensor and from the satellite itself.

Here, we provide a model of the local gravity from first principles, in which we neglect metrology uncertainties \cite{liorzou20}.
The gravitational interaction between two bodies centered on $O_1$ and $O_2$ is
\begin{equation} \label{eq_grav}
\overrightarrow{F}=-\int_{V_1}{\rm d}V_1 \int_{V_2}{\rm d}V_2 \frac{\partial V(\overrightarrow{r_1}-\overrightarrow{r_2})}{\partial r} \overrightarrow{O_1O_2},
\end{equation}
where the 3-dimensional integrals are taken over the volume of the two bodies, and $\vv{r_i} = (x_i,y_i,z_i)$ is the coordinate vector of an infinitesimal volume element of the $i$th body. 
Noting $\rho_i$ the $i$th body's density, the Newtonian potential between infinitesimal volumes
\begin{equation}
V_N(\overrightarrow{r_1}-\overrightarrow{r_2})=-\frac{G\rho_1\rho_2{\rm d}V_1{\rm d}V_2}{|\overrightarrow{r_1}-\overrightarrow{r_2}|},
\end{equation}
and the Yukawa potential of strength $\alpha$ and range $\lambda$ between infinitesimal volumes
\begin{equation}
V_Y(\overrightarrow{r_1}-\overrightarrow{r_2})=-\alpha \frac{G\rho_1\rho_2}{|\overrightarrow{r_1}-\overrightarrow{r_2}|}\exp\left(-\frac{|\overrightarrow{r_1}-\overrightarrow{r_2}|}{\lambda}\right){\rm d}V_1{\rm d}V_2.
\end{equation}

In the present case, as shown in Fig. \ref{fig_sensor}, all contributions are interactions between cylinders, either empty (test masses, electrode-bearing cylinders, shield) or full (base plate, upper clamp). For simplicity, we also assume that the vacuum system is a full cylinder. Computing the gravitational force applied to the test mass then boils down to computing the interaction between perfectly aligned cylinders (as we assumed in Sect. \ref{sect_mic}), and therefore computing the 6-dimensional integral (\ref{eq_grav}).

\ref{appG} shows that, in the limit of small displacements with which we are concerned in that article, the 6D integral (\ref{eq_grav}) can be reduced to a 1D integral depending on the geometry of the pair of cylinders. In this case, the gravitational force can be Taylor-expanded, and is dominated by a stiffness term $K_1$. The expressions given below apply both to the Newtonian ($\alpha=1$, $\lambda \rightarrow \infty$) and Yukawa forces. They give the force exerted by any one of MICROSCOPE's cylinders on a test mass.

\subsubsection{Longitudinal force}

In the limit of small displacements $\delta$ of the test mass along the cylinders' axis, the force is given by
\begin{equation} \label{eq_gravx}
{\mathcal F}_x(x_0, \delta) \approx -16\pi^2 G \rho \rho' \alpha \sum_i K_i(x_0) \delta^i,
\end{equation}
where $x_0$ is the distance between the center of the test mass and the center of the source cylinder along their longitudinal axis ($x_0=0$ when the source is either an electrode-bearing cylinder or the other test mass, but $|x_0|>0$ if the source is one cylinder of the base or the vacuum system --in which case the source and the test mass are above each other), and where the $x$ subscript corresponds to MICROSCOPE's (longitudinal) $X$-axis but is referred to as $z$ in the more conventional cylindrical coordinate system used in \ref{appG}.
The $K_i$ coefficients depend on the geometry of the test mass -- source pair as follows. If $a$ and $b$ are the inner and outer radii of the cylinder source, $2\ell$ its height and $\rho$ its density; and if $a'$ and $b'$ are the inner and outer radii of the test mass, $2L$ its height and $\rho'$ its density, then:

\begin{enumerate}
\item if the test mass is shorter than the source and they are concentric (which is the case e.g. of the pair made of the internal test mass and any electrode-bearing cylinder),
\begin{subequations}
\begin{empheq}[left={\empheqlbrace\,}]{align} 
K_0(x_0) & = 0 \\
K_1(x_0) &= \int_0^\infty \frac{W(k;a',b') W(k;a,b)}{\kappa k} {\rm e}^{-\kappa \ell} \sinh(\kappa L) {\rm d}k \\
K_2(x_0) &= 0 \\
K_3(x_0) &= \int_0^\infty \frac{\kappa}{6} \frac{W(k;a',b') W(k;a,b)}{k} {\rm e}^{-\kappa \ell} \sinh(\kappa L) {\rm d}k,
\end{empheq}
\end{subequations}
where we introduced the parameter
\begin{equation} \label{eq_kappa_param}
\kappa = \sqrt{k^2 + 1/\lambda^2}
\end{equation}
and the function
\begin{equation}
W(k;a,b) = b J_1(kb) - a J_1(ka),
\end{equation}
where $J_i$ are Bessel functions of the first kind.

\item if the test mass is longer than the source and they are concentric (which is the case of the pair made of the internal test mass as the source and the external test mass): the force is formally identical to that of the previous case, with $\ell$ and $L$ switching their roles.

\item if the test mass and the source are above each other,
\begin{subequations}
\begin{empheq}[left={\empheqlbrace\,}]{align} 
K_0(x_0) &= \frac{x_0}{|x_0|} \int_0^\infty \frac{W(k;a',b') W(k;a,b)}{\kappa^2 k} {\rm e}^{-\kappa |x_0|} \sinh(\kappa \ell) \sinh(\kappa L) {\rm d}k \\
K_1(x_0) &= -\int_0^\infty \frac{W(k;a',b') W(k;a,b)}{\kappa k} {\rm e}^{-\kappa |x_0|} \sinh(\kappa \ell) \sinh(\kappa L) {\rm d}k \\
K_2(x_0) &= \frac{x_0}{|x_0|} \int_0^\infty \frac{W(k;a',b') W(k;a,b)}{2 k} {\rm e}^{-\kappa |x_0|} \sinh(\kappa \ell) \sinh(\kappa L) {\rm d}k \\
K_3(x_0) &= -\int_0^\infty \frac{\kappa}{6k} W(k;a',b') W(k;a,b) {\rm e}^{-\kappa |x_0|} \sinh(\kappa \ell) \sinh(\kappa L) {\rm d}k.
\end{empheq}
\end{subequations}

\end{enumerate}

\subsubsection{Radial force}

Similarly, at third order in $\delta/a'$, where $\delta$ is the displacement of the test mass along a radial axis ($Y$ or $Z$), the radial force created by any one of the other cylinders is

\begin{equation} \label{eq_gravyz}
{\mathcal F}_r(x_0, \delta) \approx -2\pi^2 G \rho \rho' \alpha (K_1(x_0) \delta + K_3(x_0) \delta^3),
\end{equation}
where the $K_i$ coefficients depend on the geometry of the test mass -- source pair:

\begin{enumerate}
\item if the test mass is shorter than the source and they are nested (which is the case e.g. of the pair made of the internal test mass and any electrode-bearing cylinder),
\begin{subequations}
\begin{empheq}[left={\empheqlbrace\,}]{align} 
K_1(x_0) &= 4 \int_0^\infty \frac{k W(k;a',b') W(k;a,b)}{\kappa^2} \left[ L - \frac{{\rm e}^{-\kappa \ell}}{\kappa} \sinh(\kappa L) \cosh(\kappa |x_0|)\right] {\rm d}k \\
K_3(x_0) &= -\int_0^\infty \frac{k^3 W(k;a',b') W(k;a,b)}{\kappa^2} \left[ L - \frac{{\rm e}^{-\kappa \ell}}{\kappa} \sinh(\kappa L) \cosh(\kappa |x_0|)\right] {\rm d}k,
\end{empheq}
\end{subequations}
with $x_0 \approx 0$ in this case.

\item if the test mass is longer than the source and they are nested (which is the case of the pair made of the internal test mass as the source and the external test mass): the force is formally identical to that of the previous case, with $\ell$ and $L$ switching roles.

\item if the test mass and the source are above each other,
\begin{subequations}
\begin{empheq}[left={\empheqlbrace\,}]{align} 
K_1(x_0) &= 4 \int_0^\infty \frac{k W(k;a',b') W(k;a,b)}{\kappa^2} \frac{{\rm e}^{-\kappa |x_0|}}{\kappa} \sinh(\kappa \ell) \sinh(\kappa L) {\rm d}k \\
K_3(x_0) &= -\int_0^\infty \frac{k^3 W(k;a',b') W(k;a,b)}{\kappa^2} \frac{{\rm e}^{-\kappa |x_0|}}{\kappa} \sinh(\kappa \ell) \sinh(\kappa L) {\rm d}k
\end{empheq}
\end{subequations}
\end{enumerate}

\subsubsection{Total gravitational force}

The gravity force applied to a MICROSCOPE test mass is just the sum of the Newton and Yukawa forces created by the aforementioned instruments' parts,
\begin{align}
\vv{F}_{\rm g} &= ({\mathcal F}_{{\rm N},x} + {\mathcal F}_{{\rm Y},x}) \vv{e}_x + ({\mathcal F}_{{\rm N},r} + {\mathcal F}_{{\rm Y},r}) \vv{e}_r\\
&= \sum_j ({\mathcal F}_{{\rm N},x,j} + {\mathcal F}_{{\rm Y},x,j}) \vv{e}_x + \sum_j ({\mathcal F}_{{\rm N},r,j} + {\mathcal F}_{{\rm Y},r,j}) \vv{e}_r,
\end{align}
where the $r$ subscript stands for the $Y$ and $Z$ axes, and where the forces created by the $j$th part of the instrument ${\mathcal F}_{{\rm N},x,j}$ and ${\mathcal F}_{{\rm Y},x,j}$ are given by Eq. (\ref{eq_gravx}) and ${\mathcal F}_{{\rm N},r,j}$ and ${\mathcal F}_{{\rm Y},r,j}$ by Eq. (\ref{eq_gravyz}). 

As shown in  \ref{appG}, a first-order Taylor expansion of Eqs. (\ref{eq_gravx}) and (\ref{eq_gravyz}) is enough to precisely account for the gravitational interactions in the present article, where displacements are limited to $5~\mu{\rm m}$. This means that the local gravitation effectively acts as a stiffness on the test masses. We thus define the Newtonian and Yukawa, radial and longitudinal stiffnesses by
\begin{align}
{\mathcal F}_{{\rm N},r} &= -k_{{\rm N},r} r \\
{\mathcal F}_{{\rm N},x} &= -k_{{\rm N},x} x \\
{\mathcal F}_{{\rm Y},r} &= -k_{{\rm Y},r} r \\
{\mathcal F}_{{\rm Y},x} & = -k_{{\rm Y},x} x,
\end{align}
where $x$ and $r$ are the displacement of the test mass along the longitudinal and any radial axes of the instrument.

\begin{figure}
\begin{center}
\includegraphics[width=.7\textwidth]{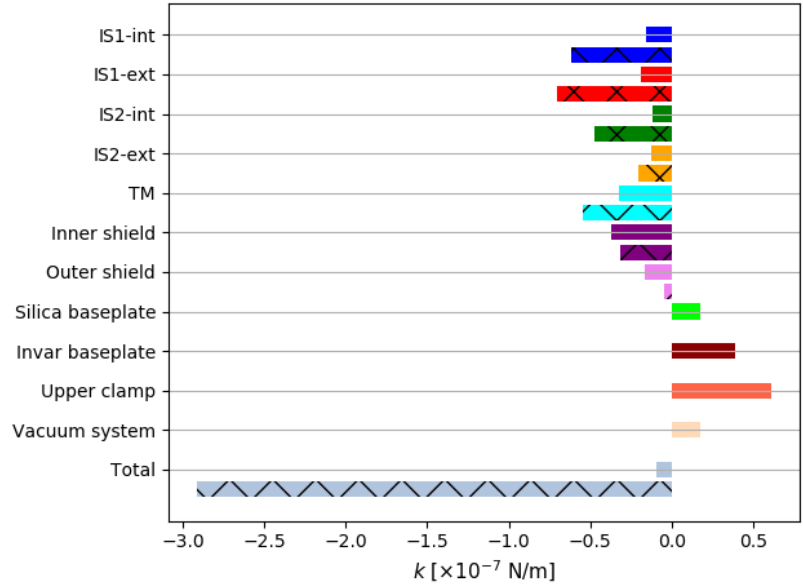}
\caption{Newtonian (plain bars) and Yukawa (hashed bars) radial stiffnesses acted by the SUEP's parts (the external test mass --TM--, the four electrode-bearing cylinders --IS1-int, IS1-ext, IS2-int, IS2-ext--, and the shielding cylinders) on its internal test mass. The Yukawa potential is set such that $(\alpha, \lambda)=(1, 0.01~{\rm m})$. The contribution of the base plates, upper clamp and vacuum system to the Yukawa interaction is too small to appear on the plot.}
\label{fig_newton}
\end{center}
\end{figure}

\paragraph{Newtonian gravity}

The plain bars of Fig. \ref{fig_newton} show the Newtonian stiffnesses from all cylinders on SUEP's internal test mass along its radial axis. The force between nested cylinders is destabilising (negative stiffness), whereas the force from the base plates, upper clamp and vacuum system stabilises the test mass, with the total radial force being destabilising. It can also be shown that the Newtonian gravitational interaction along the $X$ (longitudinal) axis acts as a stabilising stiffness.

Finally, it can be seen from the figure that the contribution from the outer shield is subdominant. Thence, those from  the other differential sensor and from the other parts of the satellite are even more subdominant, and we ignore them.

The next-to-last column of Table \ref{tab_results} lists the Newtonian gravity stiffness of the four MICROSCOPE test masses along their radial and longitudinal axes.

\paragraph{Yukawa gravity}

The hashed bars of Fig. \ref{fig_newton} show the Yukawa stiffnesses from all cylinders on SUEP's internal test mass along its radial axis, for $(\alpha, \lambda)=(1, 0.01~{\rm m})$. It can be noted that only co-axial cylinders contribute, since the base, upper clamp and vacuum system are more distant than 0.01 m from the test mass. Similarly, the closest cylinders provide most of the signal.
It can be noted that the Yukawa stiffness of the closest cylinders is larger than their Newtonian stiffness. This difference comes from the fact that with $\lambda=1~{\rm cm}$, only a restricted part of the cylinders interact, causing an effect more complex than just an exponential decay proportional to the Newtonian stiffness.

\begin{figure}
\begin{center}
\includegraphics[width=.45\textwidth]{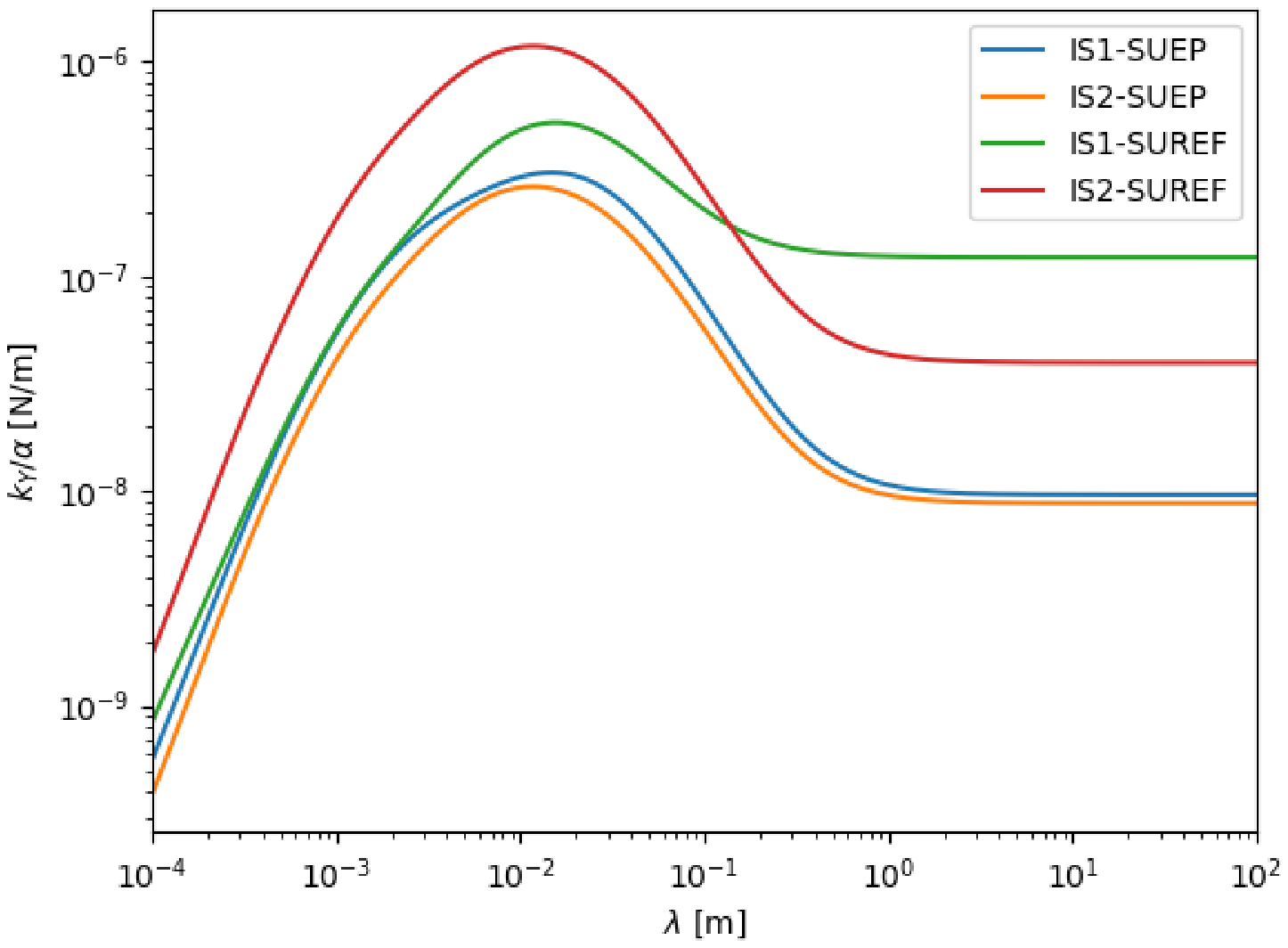}
\includegraphics[width=.45\textwidth]{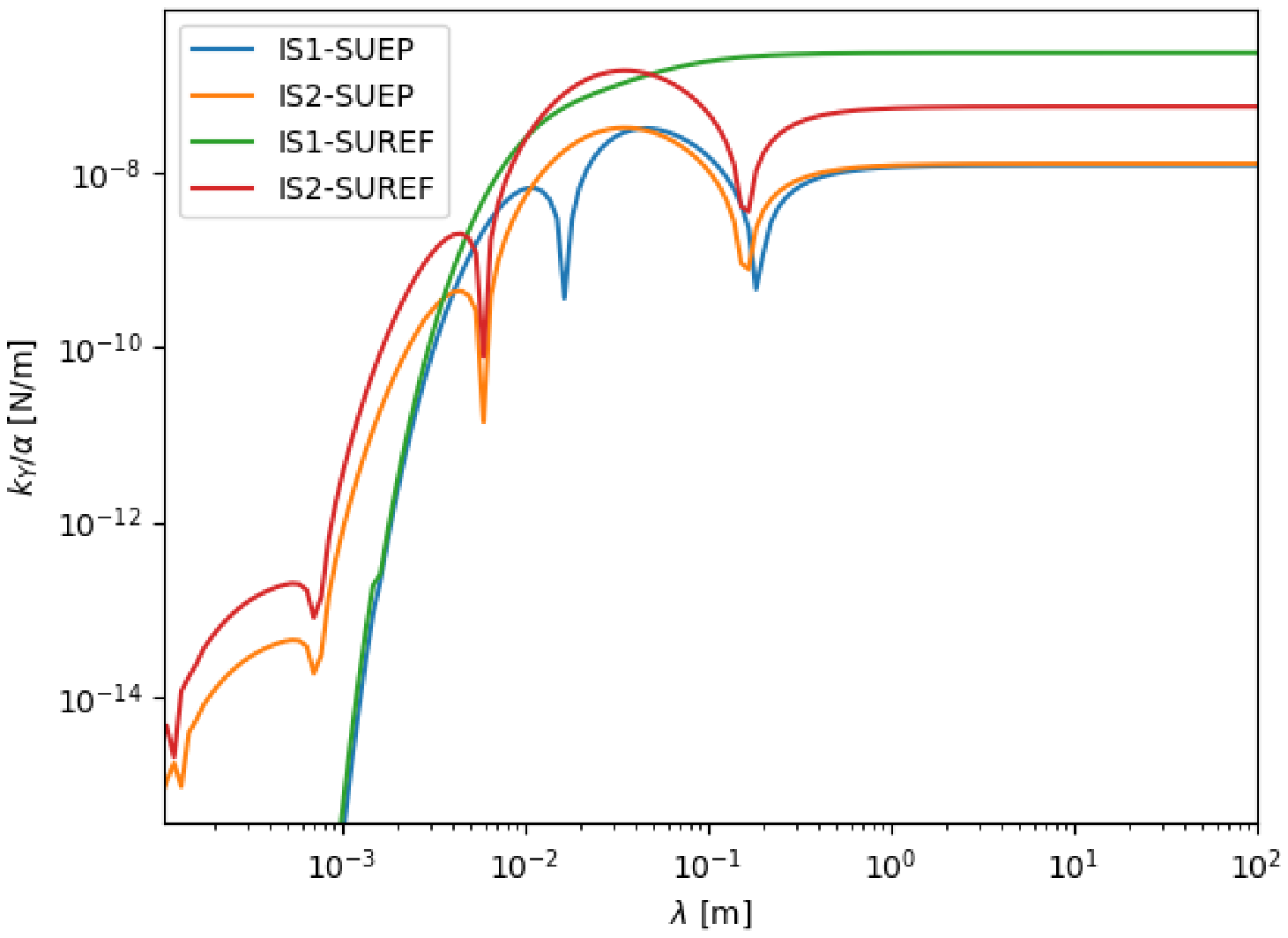}
\caption{Yukawa stiffness (normalised by $\alpha$) for the four MICROSCOPE test masses, on the radial (left) and longitudinal (right) axes, as a function of Yukawa's range.}
\label{fig_klambda}
\end{center}
\end{figure}

Fig. \ref{fig_klambda} shows Yukawa gravity's stiffness as a function of Yukawa's range $\lambda$ for all MICROSCOPE test masses, along their radial (left panel) and longitudinal (right panel) axes. Starting from the smaller $\lambda$ reachable (linked to the distance between a test mass and its closest cylinder), the radial stiffness increases steadily as more and more co-axial cylinders are within reach of $\lambda$ and contribute to the gravity signal. The stiffness peaks around $\lambda \approx 0.01~{\rm m}$, where the base and upper cylinders start to contribute but with an opposite sign stiffness, thereby decreasing it until the Newtonian regime is reached when $\lambda$ becomes larger than the sensor's largest scale. The longitudinal stiffness shows a similar behaviour, though it changes sign while more and more cylinders contribute to the signal.

Comparing Fig. \ref{fig_klambda} with Table \ref{tab_results}, it is clear that the gravity stiffness (and therefore, signal) is largely subdominant. We put it to test in  Sect. \ref{sect_analysis}.

\subsection{Summary: measured acceleration}

Taking all the forces above into account, the acceleration (\ref{eq_main2}) of a test mass measured along the $i$th axis during a stiffness characterisation session is
\begin{multline} \label{eq_completeAcc}
\Gamma_i(t) = b_{\epsilon,i} + a_{p,i} + a_{r,i} + \frac{m \omega^2 + k_{\epsilon,i} + k_{N,i}}{m} x_{i0} \sin(\omega t + \psi) \\
+ \frac{k_{w,i}}{m}x_{i0} \sin(\omega t + \psi - \phi) 
+ \frac{\lambda_{w,i}}{m} \omega x_{i0} \cos(\omega t + \psi) 
+ \frac{k_{Y,i}(\alpha, \lambda)}{m_j}x_{i0} \sin(\omega t + \psi),
\end{multline}
where we singled out the Yukawa gravity contribution and made its stiffness' dependence on $(\alpha, \lambda)$ explicit, since it is this very dependence that we aim to constrain in the remainder of this article. The first three terms of the equation comprise the effect of the measurement bias, radiation pressure and photometric effect (all three acting as a constant bias) and the fourth term encompasses the in-phase stiffnesses (excitation, electrostatic and Newtonian gravity). The last three terms give the effect of the gold wire (internal and viscous damping), and the Yukawa gravity contribution. 

\section{Data analysis} \label{sect_analysis}


This section presents least-squares estimates of the parameters introduced in the previous section. We use the acceleration data introduced in subsection \ref{ssect_measurement} to estimate two components of the stiffness (that of the gold wire $k_w$ and a linear combination $k_0$ of the electrostatic and gravitational stiffness), the quality factor of the gold wire, and a velocity-dependent coefficient for each axis of each sensor (Eq. \ref{eq_recast} below). Then, in subsection \ref{sssect_results}, we subtract theoretical models of the electrostatic and Newtonian gravity stiffness from the estimated $k_0$; any residual should come either from a Yukawa stiffness or from unaccounted for contributors.
We perform the exercice in the two electrical configurations (HRM and FRM) summarised in \ref{app_hrmfrm}.

\subsection{Measurement equation} \label{ssect_anaIntro}

The measurement equation (\ref{eq_completeAcc}) could be used in its original form to extract the unknown parameters from the data and simultaneously constrain the Yukawa interaction's parameters. However, since the Yukawa contribution is expected to be at most of the order of the Newtonian contribution, which is itself largely less than the electrostatic stiffness, its parameters have a small constraining power on the data, and we find more suited to first estimate an overall stiffness, from which we can eventually extract the $(\alpha, \lambda)$ parameters.

Moreover, Eq. (\ref{eq_completeAcc}) requires the estimation of two phases. The first one, $\psi$, is that of the excitation signal and can be estimated {\it a priori} by fitting the position data, then used as a known parameter in the following analysis. The second, $\phi$, is the phase-offset induced by the gold wire's internal damping. 
Instead of trying to estimate it from the data (which may be difficult given the 4 Hz sampling of data, when assuming that the quality factor of the wire is in the range $Q\approx 1-100$, corresponding to a time offset less than 2 s), we recast Eq. (\ref{eq_completeAcc}) as
\begin{equation} \label{eq_recast}
\Gamma(t) = b + \left(\kappa_0 + \kappa_w \cos\phi \right) \sin(\omega t + \psi) - \left(\kappa_w \sin\phi  - \kappa_\lambda \right) \cos(\omega t + \psi),
\end{equation}
where $b \equiv b_{\epsilon} + a_p + a_r$, $\kappa_0 \equiv x_0 (m \omega^2 + k_\epsilon + k_N + k_Y) / m$, $\kappa_w \equiv x_0 k_w / m$ and $\kappa_\lambda \equiv x_0 \lambda_w \omega / m$, and we dropped the $i$ subscript for simplicity. 

Five parameters are left for estimation: $b$, $\kappa_0$, $\kappa_w$, $\kappa_\lambda$ and $\phi=1/Q$. It is however clear that fitting Eq. (\ref{eq_recast}) will provide only three independent constraints. If estimating $b$ will be easy, the other parameters will remain degenerate unless we can use some prior knowledge. We show in Sect. \ref{ssect_kQ} that we can obtain an independent estimate of the $\kappa_w/Q = \kappa_w \phi$ combination.

\subsection{Gold wire's $k_w/Q$ ratio} \label{ssect_kQ}

\begin{figure}
\begin{center}
\includegraphics[width=.5\textwidth]{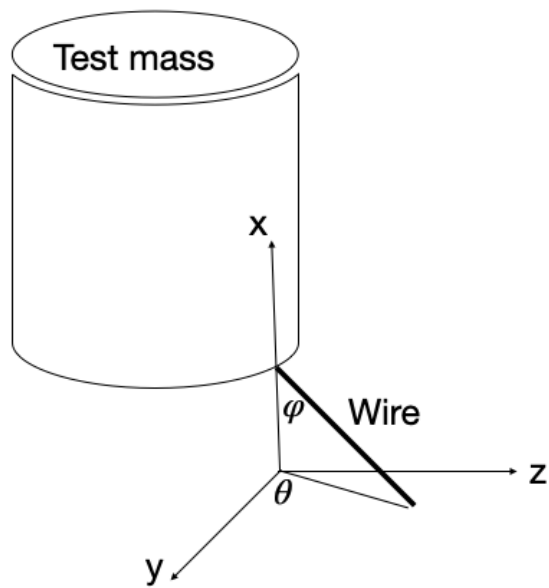}
\caption{Gold wire and test mass geometry.}
\label{fig_wiregeometry}
\end{center}
\end{figure}

\begin{figure}
\begin{center}
\includegraphics[width=.7\textwidth]{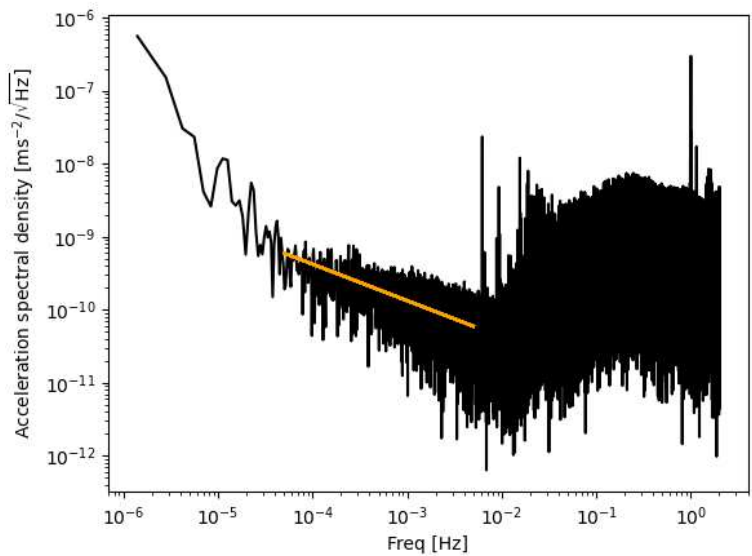}
\caption{Typical (square-root) spectral density of the acceleration measured along the $X$-axis of IS1-SUEP. The $f^{-1/2}$ low-frequency part originates from thermal dissipation in the wire (Eq. \ref{eq_goldwirePSD}), and the high-frequency $f^2$ increase is due to the capacitive detector's noise \cite{touboul09}. The orange line is the best fit of the low-frequency part.}
\label{fig_spectrumX}
\end{center}
\end{figure}

As shown in Sect. \ref{ssect_gw}, fitting the low-frequency part of the spectral density of the acceleration measured along a given axis can provide an estimate of the ratio $k_w/Q$ for this axis (from Eq. \ref{eq_goldwirePSD}) once temperature data are available (which is the case for all measurement sessions). 
Performing this task for the three linear axes, we can get an estimate of the gold wire stiffness along each axis, and the orientation of the force due to the wire. This force is presumably collinear with the wire, although the glue clamping process may complexify it. Noting $\varphi$ the angle between the force and the test mass longitudinal axis ($X$-axis), and $\theta$ the angle between the $Y$-axis and the projection of the wire on the ($y,z$) plane (Fig. \ref{fig_wiregeometry}), the three stiffnesses that can be measured are
\begin{eqnarray}
k_{w,x} & = & |k_w| \cos\varphi \\
k_{w,y} & = & |k_w| \sin\varphi \cos\theta \\
k_{w,z} & = & |k_w| \sin\varphi \sin\theta
\end{eqnarray}
from which we can recover the modulus of the stiffness and the orientation of the wire. Fig. \ref{fig_spectrumX} shows the fit corresponding to IS1-SUEP's $X$-axis from the session used to estimate the WEP in Ref. \cite{touboul17}. Values obtained for the internal sensor of both SU are given in Table \ref{tab_egw}. We checked that estimates from different sessions are consistent. 
Note that rigorously, since the drag-free system is controlled by the external sensor, fitting the internal sensor's spectral density only provides information about the sum of the $k_w/mQ$ ratios of both sensors (where $m$ is their mass). Nevertheless, under the assumptions that their masses are similar (which is enough given the goals of this article), that their wires have similar $k_w/Q$ ratios, and that their spectral density are uncorrelated, fitting the internal sensor's spectral density indeed provides a constraint on each sensor's wire's $k_w/Q$ ratio.

\begin{table}
\caption{Gold wire stiffness [$10^{-3}$Nm$^{-1}$] and orientation of the force [deg] estimated from long measurement sessions' acceleration noise spectral density.}
\begin{center}
\begin{tabular}{ccccccc}
\hline
& $k_{w,x}/Q$ & $k_{w,y}/Q$ & $k_{w,z}/Q$ & $k_w/Q$ & $\varphi$ & $\theta$ \\
\hline
IS1-SUEP & $1.14\pm0.20$ & $0.31\pm0.05$ & $1.26\pm0.20$ & $1.73\pm0.10$ & $48\pm4$ & $70\pm25$ \\
IS1-SUREF & $0.07\pm0.01$ & $0.22\pm 0.10$ & $0.66\pm0.06$ & $0.70\pm0.08$ & $84\pm1$ & $66\pm30$ \\
\hline
\end{tabular}
\end{center}\label{tab_egw}
\label{default}
\end{table}%

Willemenot \& Touboul \cite{willemenot00} used a torsion pendulum to characterise a gold wire similar to those used by MICROSCOPE. Assuming that the wire is deformed perpendicular to its principal axis (i.e. in flexion), they give a convenient scaling to quantify the wire's stiffness
\begin{equation}
k_w = 2.91\times10^{-5} \left(\frac{r_w}{3.75~\mu{\rm m}}\right)^4 \left(\frac{1.7~{\rm cm}}{l_w}\right)^3 \left(\frac{E}{7.85\times10^{10}~{\rm Nm}^2}\right),
\end{equation}
where $r_w$ is the radius of the wire, $l_w$ its length and $E$ its Young modulus. Using MICROSCOPE's gold wires' characteristics ($r_w=3.5~\mu{\rm m}$, $l_w=2.5~{\rm cm}$ and $E~=~7.85~\times~10^{10}~{\rm Nm}^2$), we expect $k_w \approx 9\times10^{-6}$.
Combined with a quality factor $Q\approx 100$ as measured in Ref. \cite{willemenot00}, this scaling provides $k_w/Q \approx 10^{-7}~{\rm N/m}$, in flagrant contradiction with the values estimated from flight data (Table \ref{tab_egw}).

Two explanations can be proposed: (i) the wire does not behave as shown in Ref. \cite{willemenot00} or (ii) its quality factor is much lower than expected. 
In the former explanation, the wire may work in compression (i.e. it is deformed along its principal axis), which potentially increases its stiffness. In the latter, the mounting process (wires being glued to the test masses) may decrease the overall quality factor; differences between the glue points in MICROSCOPE and in Ref. \cite{willemenot00} may explain a significant difference of quality factor.

Assuming that the electrostatic model of the instrument is correct and that the measured stiffness is dominated by the electrostatic stiffness hints at a low quality factor $Q\approx 1$. 
Note however that, even if the quality factor is really that low, MICROSCOPE's main results (the test of the WEP) depend on the $k_w/Q$ ratio, and are thus unaffected by the current analysis.

\subsection{Stiffness estimation}  \label{ssect_kestimation}

We now come back to Eq. (\ref{eq_recast}), with the aim to estimate the model parameters for the four sensors, starting with radial axes. 

\subsubsection{Radial axes ($Y$ and $Z$)}

The following assumptions allow us to break the degeneracy between the parameters mentioned in Sect. \ref{ssect_anaIntro}:
\begin{itemize}
\item for a given sensor and a given axis $j$, the gold wire's ratio $k_{w,j}/Q$ is independent of the electrical configuration (HRM or FRM), and can be estimated as shown in Sect. \ref{ssect_kQ} for the internal sensors. We further assume that the mounting of gold wires is general enough to assume that the external sensors' $k_{w,j}/Q$ ratio is the same as that of the internal sensor \cite{liorzou20}.
\item the ratio $k_{w,j}/Q$ varies from one axis to another, but the quality factor $Q$ is a true constant for a given sensor. In other words, since $k_{w,j}$ and $Q$ are degenerate, we assume that only the stiffness depends on the direction.
\item by cylindrical symmetry, the total stiffness of the radial axis $j$ ($j=y,z$) $k_{0,j} ~= ~m ~\omega^2 ~+ ~k_{\epsilon,j} ~+ ~k_N ~+ ~k_Y$ is independent of the axis, and depends on the electrical configuration only through the electrostatic stiffness $k_\epsilon$. This assumption is reasonable given the estimated metrology uncertainties \cite{liorzou20}.
\end{itemize}

Denoting $\hat{\chi}_y=k_{w,y}/Q$ and $\hat{\chi_z}=k_{w,z}/Q$ the radial gold wire's ratios estimated in Table \ref{tab_egw}, and combining constraints from the model (\ref{eq_recast}), where we add the subscripts `F' and `H' for measurements in FRM and HRM modes, we obtain the following system of equations for a given sensor:
\begin{subequations}
\begin{empheq}[left={\empheqlbrace\,}]{align} 
\label{eq_1}
\hat{a}_{0yF} & = \kappa_{0rF} + \kappa_{wy} \cos\phi \\
\label{eq_3}
\hat{a}_{0yH} & = \kappa_{0rH} + \kappa_{wy} \cos\phi \\
\label{eq_5}
\hat{a}_{0zF} & = \kappa_{0rF} + \kappa_{wz} \cos\phi \\
\label{eq_7}
\hat{a}_{0zH} & = \kappa_{0rH} + \kappa_{wz} \cos\phi \\
\label{eq_2}
\hat{a}_{wyF} & = -\kappa_{wy} \sin\phi + \kappa_{\lambda yF} \\
\label{eq_4}
\hat{a}_{wyH} & = -\kappa_{wy} \sin\phi  + \kappa_{\lambda yH}  \\
\label{eq_6}
\hat{a}_{wzF} & = -\kappa_{wz} \sin\phi + \kappa_{\lambda zF} \\
\label{eq_8}
\hat{a}_{wzH} & = -\kappa_{wz} \sin\phi  + \kappa_{\lambda zH}  \\
\label{eq_9}
\hat{\chi}_y & = \frac{k_{wy}}{Q} \\
\label{eq_10}
\hat{\chi}_z & = \frac{k_{wz}}{Q},
\end{empheq}
\end{subequations}
where we recall that $\phi=1/Q$, $\kappa_{wj} = x_0 k_{wj} / m$, and similarly for $\kappa_{0r}$ (with the subscript $r=y,z$), $\kappa_\lambda$ and the $\hat{a}_0$ and $\hat{a}_w$ coefficients are the estimates of the sine and cosine coefficients of Eq. (\ref{eq_recast}).

On the one hand, Eqs. (\ref{eq_2}-\ref{eq_8}) trivially give the velocity-dependent terms as a functions of the unknown $Q$ and estimated $\chi_j$ and $a_{wj}$. On the other hand, Eqs. (\ref{eq_1}-\ref{eq_7}), (\ref{eq_9}-\ref{eq_10}) can be combined to give
\begin{equation}
2(\hat{\chi}_y - \hat{\chi}_z) \frac{x_0}{m} Q \cos\left(\frac{1}{Q}\right) = \hat{a}_{0yF} + \hat{a}_{0yH} - \hat{a}_{wzF} - \hat{a}_{wzH},
\end{equation}
thus providing the equation
\begin{equation}
x \cos\left(\frac{1}{x}\right) - \xi = 0
\end{equation}
of which $Q$ is a root, where $\xi$ is defined through parameters estimated from Table \ref{tab_egw} and fitting Eq. (\ref{eq_recast}) for the sensor's two radial axes in each electrical configuration.

Once $Q$ is estimated, Eqs. (\ref{eq_1}-\ref{eq_7}) readily provide $\kappa_{0rF}$ and $\kappa_{0rH}$. Actually, they give two estimates of each, which we checked to be consistent.

\subsubsection{Longitudinal axis $X$}

Under the same assumptions, it is then straightforward to estimate the $X$-axis stiffness from Eq. (\ref{eq_recast}) and Table \ref{tab_egw}, for a given electrical configuration (that we do not make explicit in the equations below for simplicity):
\begin{subequations}
\begin{empheq}[left={\empheqlbrace\,}]{align} 
\kappa_{0x} & = \hat{a}_{0x} - \frac{x_0}{m} \hat{\chi}_x Q\cos\left(\frac{1}{Q}\right) \\
\kappa_{\lambda x} & = \hat{a}_{wx} + \frac{x_0}{m} \hat{\chi}_x Q\sin\left(\frac{1}{Q}\right).
\end{empheq}
\end{subequations}

\subsubsection{Results} \label{sssect_results}

Our results are listed in Table \ref{tab_results} for each sensor, in their two electrical configurations. The left group's four columns show the instrumental parameters: total in-phase stiffness $k_0$, gold wire stiffness $k_w$ and quality factor $Q$, and velocity-dependent coefficient $\lambda_w$. The next two columns give the theoretical electrostatic stiffness $k_{\epsilon, {\rm th}}$ and the Newtonian gravity stiffness. The last column lists the difference between the theoretical and the estimated electrostatic stiffness $\Delta k = \hat{k}_0 - k_N - m \omega^2 - k_{\epsilon, {\rm th}}$. Error bars give $1\sigma$ uncertainties.

\begin{figure}
\begin{center}
\includegraphics[width=.7\textwidth]{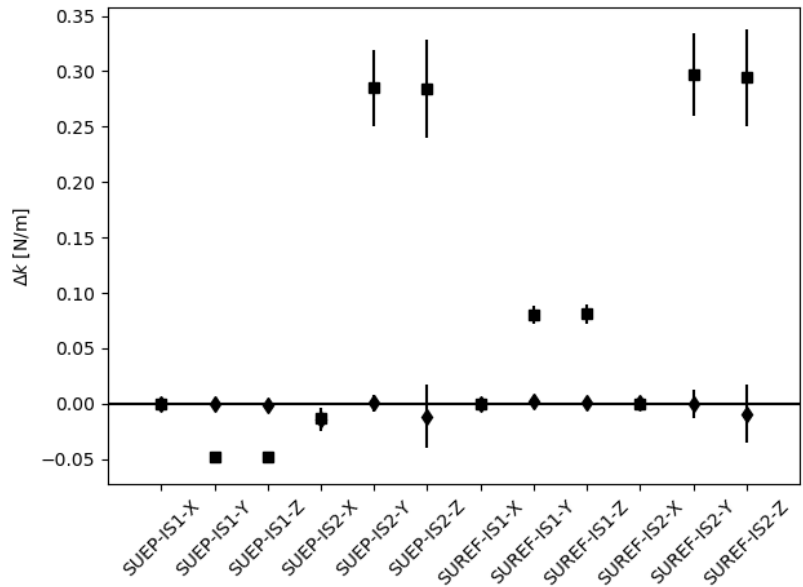}
\caption{Difference between theoretical electrostatic stiffness and measured total in-phase stiffnesses corrected for the excitation and Newtonian gravity stiffnesses, $\Delta k = \hat{k}_0 - k_N - m \omega^2 - k_{\epsilon, {\rm th}}$, for all axes (longitudinal and radial) of each sensor, in the HRM (diamonds) and FRM (squares) electric configurations.}
\label{fig_deltak}
\end{center}
\end{figure}

The electrostatic stiffness estimated in the HRM electrical configuration is consistent with the theoretical one for most sensors and axes, with at most a $\approx 2\sigma$ discrepancy. However, we note a significant difference in the FRM configuration (Fig. \ref{fig_deltak}). Being dependent on the electrical configuration, this discrepancy hints at the existence of an electric potential-dependent additional stiffness completely degenerate with the electrostatic one. Patch effects could be at its origin: the discrepancy being significant for higher voltages is indeed consistent with the voltage-dependence of patch effects' stiffness. Disentangling this puzzle would require modeling patch effects in MICROSCOPE's sensors. As this goes far beyond the scope of this paper, and since our experiment is not competitive with other short-ranged forces searches (as we discuss below), we let the question of this discrepancy open. In the remainder of this paper, we thence use HRM measurements only.

The gold wire's quality factor is lower than could be expected from Ref. \cite{willemenot00}. Nevertheless, $Q$ being close to 1 is consistent with our discussion in Sect. \ref{ssect_kQ}. Correspondingly, the gold wires' stiffness is small and negligible compared to the electrostatic stiffness of the radial axes, as was assumed by Ref. \cite{chhuncqg5}; however,  those stiffnesses remain significantly degenerate in the longitudinal axis, giving a total stiffness similar to that estimated in Ref. \cite{chhuncqg5}.
Moreover, this degeneracy, that also exists for the radial axes, explains that the total errors are dominated by the uncertainty on the gold wire's stiffness ($k_0$ and $k_w$ are degenerate in the amplitude of Eq. (\ref{eq_recast})'s sine, meaning that $\hat{k}_0$ errors actually come from those on $k_w/Q$).

\begin{landscape}
\begin{table}
\caption{Estimated model parameters. The left group's four columns show the instrumental parameters: total in-phase stiffness $k_0$, gold wire stiffness $k_w$ and quality factor $Q$, and velocity-dependent coefficient $\lambda_w$. The next two columns give the theoretical electrostatic stiffness $k_{\epsilon, {\rm th}}$ and the Newtonian gravity stiffness. The last column lists the difference between the theoretical and the estimated electrostatic stiffness $\Delta k = \hat{k}_0 - k_N - m \omega^2 - k_{\epsilon, {\rm th}}$. Error bars give $1\sigma$ uncertainties.
} 
\small
\begin{center}
\begin{tabular}{cc|cccc||cc||c}
\hline
Sensor & Axis (mode) & $\hat{k}_0$ & $\hat{k}_w$ & $\hat{Q}$ & $\hat{\lambda}_w$ & $k_{\epsilon, {\rm th}}$ & $k_N$ & $\Delta k$ \\
 & &  [$\times10^{-2}~{\rm N/m}$] &  [$\times10^{-2}~{\rm N/m}$] & &  [$\times10^{-2}~{\rm Ns/m}$] &  [$\times10^{-2}~{\rm N/m}$] & [$\times 10^{-8}$ N/m] &  [$\times10^{-2}~{\rm N/m}$] \\
\hline
SUEP & $X$ (HRM) & -0.00$\pm$0.12& 0.16$\pm$0.08 & 1.5$\pm$0.7 & 4.23$\pm$3.82 & 0.00$\pm$0.01 & 1.22 & -0.021$\pm$0.119 \\
IS1 & $Y$ (HRM) & -1.55$\pm$0.06& 0.04$\pm$0.08 & 1.5$\pm$0.7 & 1.87$\pm$1.87 & -1.57$\pm$0.05 & -0.96 & 0.009$\pm$0.081 \\
 & $Z$ (HRM) & -1.65$\pm$0.26& 0.19$\pm$0.31 & 1.5$\pm$0.7 & 5.50$\pm$7.49 & -1.57$\pm$0.05 & -0.96 & -0.095$\pm$0.260 \\
 & $X$ (FRM) & 0.04$\pm$0.12& 0.16$\pm$0.08 & 1.5$\pm$0.7 & 2.88$\pm$3.89 & 0.00$\pm$0.01 & 1.22 & 0.024$\pm$0.120 \\
 & $Y$ (FRM) & -18.85$\pm$0.06& 0.04$\pm$0.08 & 1.5$\pm$0.7 & 1.35$\pm$1.89 & -14.08$\pm$0.46 & -0.96 & -4.795$\pm$0.460 \\
 & $Z$ (FRM) & -18.87$\pm$0.26& 0.19$\pm$0.31 & 1.5$\pm$0.7 & 4.94$\pm$7.53 & -14.08$\pm$0.46 & -0.96 & -4.809$\pm$0.523 \\
\hline
SUEP & $X$ (HRM) & -1.35$\pm$1.01& 1.41$\pm$1.01 & 12.8$\pm$8.9 & 5.60$\pm$2.93 & 0.00$\pm$0.01 & 1.27 & -1.366$\pm$1.015 \\
IS2 & $Y$ (HRM) & -6.96$\pm$0.73& 0.40$\pm$0.73 & 13.4$\pm$9.5 & 5.03$\pm$2.38 & -7.01$\pm$0.23 & -0.88 & 0.039$\pm$0.760 \\
 & $Z$ (HRM) & -8.20$\pm$2.94& 1.74$\pm$2.95 & 13.4$\pm$9.5 & 9.65$\pm$9.53 & -7.01$\pm$0.23 & -0.88 & -1.201$\pm$2.949 \\
 & $X$ (FRM) & -1.32$\pm$1.02& 1.41$\pm$1.01 & 12.8$\pm$8.9 & 5.96$\pm$2.97 & 0.00$\pm$0.01 & 1.27 & -1.337$\pm$1.015 \\
 & $Y$ (FRM) & -78.47$\pm$0.68& 0.38$\pm$0.69 & 12.7$\pm$8.6 & 2.25$\pm$2.39 & -107.02$\pm$3.37 & -0.88 & 28.538$\pm$3.440 \\
 & $Z$ (FRM) & -78.56$\pm$2.77& 1.66$\pm$2.78 & 12.7$\pm$8.6 & 7.03$\pm$9.53 & -107.03$\pm$3.37 & -0.88 & 28.456$\pm$4.366 \\
\hline
\hline
SUREF & $X$ (HRM) & 0.06$\pm$0.09& 0.02$\pm$0.01 & 2.3$\pm$1.3 & -0.52$\pm$4.35 & 0.00$\pm$0.01 & 23.65 & 0.041$\pm$0.092 \\
IS1 & $Y$ (HRM) & -1.58$\pm$0.08& 0.05$\pm$0.08 & 2.5$\pm$1.3 & 1.60$\pm$1.33 & -1.81$\pm$0.06 & -12.32 & 0.209$\pm$0.095 \\
 & $Z$ (HRM) & -1.71$\pm$0.12& 0.18$\pm$0.12 & 2.5$\pm$1.3 & 3.80$\pm$1.40 & -1.81$\pm$0.06 & -12.32 & 0.082$\pm$0.133 \\
 & $X$ (FRM) & 0.06$\pm$0.09& 0.02$\pm$0.01 & 2.3$\pm$1.3 & -0.92$\pm$4.51 & 0.00$\pm$0.01 & 23.65 & 0.042$\pm$0.095 \\
 & $Y$ (FRM) & -19.25$\pm$0.07& 0.05$\pm$0.08 & 2.5$\pm$1.3 & 5.72$\pm$1.33 & -27.31$\pm$0.86 & -12.32 & 8.047$\pm$0.862 \\
 & $Z$ (FRM) & -19.16$\pm$0.12& 0.18$\pm$0.12 & 2.5$\pm$1.3 & 7.79$\pm$1.41 & -27.31$\pm$0.86 & -12.32 & 8.134$\pm$0.866 \\
\hline
SUREF & $X$ (HRM) & 0.20$\pm$0.35& 0.20$\pm$0.17 & 28.9$\pm$20.9 & 0.85$\pm$14.57 & 0.00$\pm$0.01 & 5.72 & 0.144$\pm$0.347 \\
IS2 & $Y$ (HRM) & -8.91$\pm$1.09& 0.68$\pm$1.09 & 33.9$\pm$19.2 & 5.81$\pm$1.44 & -9.09$\pm$0.29 & -3.98 & 0.111$\pm$1.126 \\
 & $Z$ (HRM) & -9.56$\pm$1.69& 2.38$\pm$1.68 & 33.9$\pm$19.2 & 9.34$\pm$2.00 & -9.08$\pm$0.29 & -3.98 & -0.540$\pm$1.710 \\
 & $X$ (FRM) & 0.15$\pm$0.34& 0.20$\pm$0.17 & 28.9$\pm$20.9 & -2.96$\pm$14.45 & 0.00$\pm$0.01 & 5.72 & 0.089$\pm$0.345 \\
 & $Y$ (FRM) & -80.24$\pm$1.04& 0.65$\pm$1.04 & 32.4$\pm$19.3 & 21.79$\pm$1.45 & -110.17$\pm$3.47 & -3.98 & 29.877$\pm$3.629 \\
 & $Z$ (FRM) & -80.09$\pm$1.66& 2.27$\pm$1.66 & 32.4$\pm$19.3 & 24.64$\pm$1.88 & -110.13$\pm$3.47 & -3.98 & 29.988$\pm$3.847 \\
\hline
\hline
\end{tabular}
\end{center}\label{tab_results}
\label{default}
\end{table}%

\end{landscape}

\section{Constraints on short-ranged Yukawa deviation} \label{sect_results}

In the previous section, we invoked patch effects to account for the non-zero difference between theoretical electrostatic stiffness and measured total in-phase stiffness corrected for the excitation and Newtonian gravity stiffnesses, $\Delta k = \hat{k}_0 - k_N - m \omega^2 - k_{\epsilon, {\rm th}}$. Actually, $\Delta k$ also contains the putative Yukawa potential that we aim to constrain in this paper.

\begin{figure}
\begin{center}
\includegraphics[width=.7\textwidth]{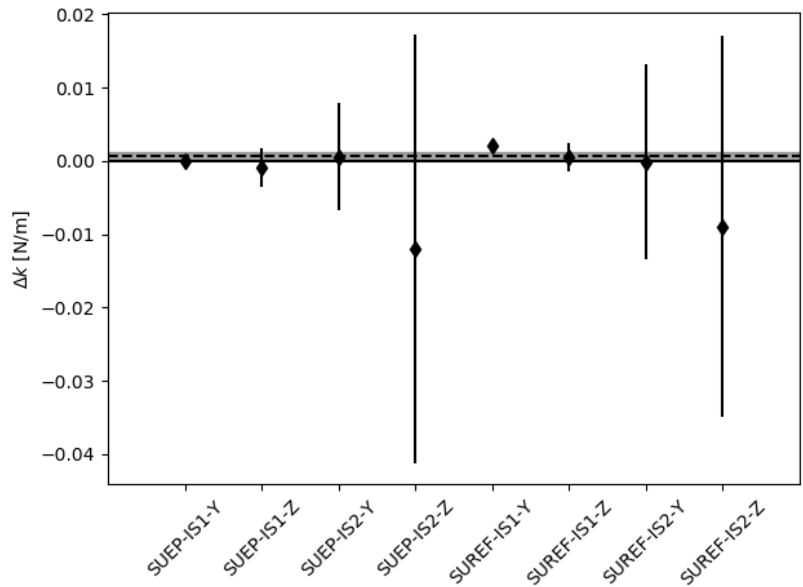}
\caption{Difference between theoretical electrostatic stiffnesses and measured total in-phase stiffness corrected for the excitation and Newtonian gravity stiffnesses, $\Delta k = \hat{k}_0 - k_N - m \omega^2 - k_{\epsilon, {\rm th}}$, for the radial axes of each sensor, in the HRM electric configuration. The dashed line is the $\Delta k$ weighted average and the grey area shows its  $1\sigma$ weighted uncertainty.}
\label{fig_deltakHRM}
\end{center}
\end{figure}

Given the obvious dependence of $\Delta k$ on the electric configuration, which cannot be explained by a Yukawa-like gravity interaction, we exclude the obviously biased FRM measurements from our analysis below. Furthermore, as shown in Fig. \ref{fig_klambda}, a Yukawa potential has a stronger signature on the radial axes than on the longitudinal one. Therefore, we use only the stiffness estimated on the radial axes in the HRM configuration to infer constraints on the Yukawa interaction.
Fig. \ref{fig_deltakHRM} shows the corresponding $\Delta k$ estimate, together with their weighted average and $1\sigma$ uncertainty (dashed line and grey area), $\langle \Delta k \rangle = (7.1\pm6.0)\times10^{-4}~{\rm N/m}$.
It can be noted that SUEP and SUREF have similar behaviours. This is expected since they are identical --up to their external test mass and small machining errors.

\begin{figure}
\begin{center}
\includegraphics[width=.7\textwidth]{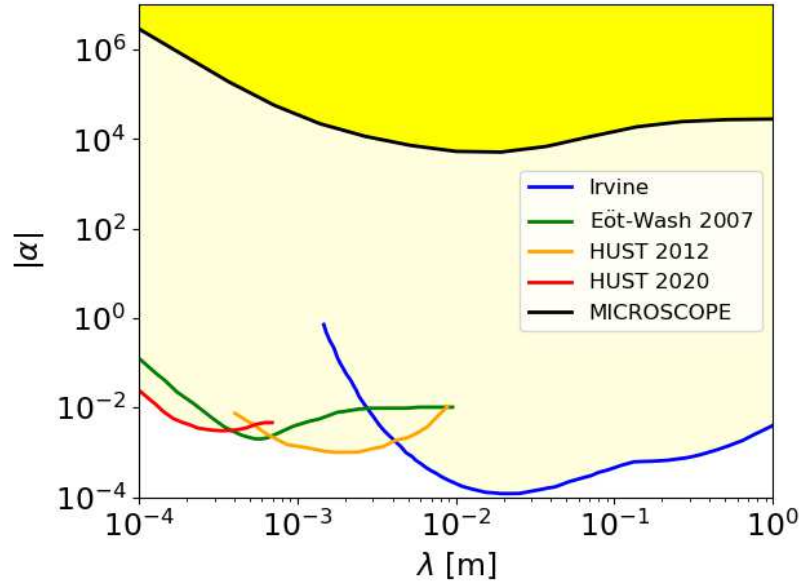}
\caption{95\% confidence contour for a Yukawa potential. The light yellow area shows the excluded region by various experiments: Irvine \cite{hoskins85}, E\"ot-Wash 2007 \cite{kapner07}, HUST 2012 \cite{yang12}, HUST 2020 \cite{tan20}, and the yellow area shows the region excluded by the current work.}
\label{fig_yc}
\end{center}
\end{figure}

The marginal offset from 0 is surely due to unaccounted for patch effects and a possible suboptimal calibration of our electrostatic model. However, as error bars are largely dominated by gold wires, and are significantly larger than the remaining bias, we use this estimation of $\langle \Delta k \rangle$ to infer the 95\% (2~$\sigma$) upper bound on the Yukawa potential in Fig. \ref{fig_yc}, noting that a positive $\langle \Delta k \rangle$ corresponds to a negative $\alpha$.
Note that since our estimated $\Delta k$ is consistent with 0, we merely consider that a Yukawa interaction can be present within the error bars; we do not claim that it explains $\Delta k$'s slight offset from 0.

The curves in the lower part of Fig. \ref{fig_yc} show the current best upper bounds on a Yukawa potential, inferred from dedicated torsion balance experiments \cite{hoskins85, kapner07, yang12, tan20}. 
Note that the E\"ot-Wash group recently updated its constraints \cite{lee20}; however, since they have been improved below the ranges of $\lambda$ considered here, we do not show them in Fig.~\ref{fig_yc}.
Our constraints are clearly poor compared to the state of the art. It would have been surprising otherwise, since MICROSCOPE was not designed to look for short-range deviations from Newtonian gravity. However, our results suggest that thanks to its non-trivial geometry, an experiment looking like MICROSCOPE, if highly optimised, may allow for new constraints of gravity through the measurement of the interaction between several bodies.

\section{Conclusion} \label{sect_conclusion}

We used in-flight technical measurements aimed to characterise MICROSCOPE's instrument to search for short-ranged Yukawa deviations from Newtonian gravity. MICROSCOPE not being designed for this task, this article serves as a proposal for a new experimental concept in the search of small-scale modifications of gravitation, as well as a first proof of concept. The analysis is based on the estimation of the stiffness of the force underwent by MICROSCOPE's test masses as they are set in motion in their cage. 

We listed all forces possibly intervening in the measurement, and computed the total stiffness. We found that estimation uncertainties are dominated by those coming from the gold wires' stiffness and quality factor (those wires being used to control the potential of the test masses).
As the electrostatic stiffness is expected to dominate over other stiffnesses, we compared it with the estimated total stiffness. We found a non-zero difference depending on the instrument's electric configuration, hinting at unaccounted for patch field forces. Due to the complexity of their modeling, we removed those measurements with a significant bias from our inference of the Yukawa potential parameters.

Not surprisingly, our constraints on the Yukawa potential parameter space ($\alpha$, $\lambda$) are not competitive with the published ones, obtained with dedicated laboratory tests. We find $|\alpha| < 10^4-10^6$ for $10^{-4}~{\rm m} \leqslant \lambda \leqslant 1~{\rm m}$, eight orders of magnitude above the best current upper bounds. Nevertheless, our work can be the starting point for optimisations to be implemented in the proposed MICROSCOPE's follow up. The gold wire should be replaced by a contactless charge control management, as envisioned for LISA \cite{sumner09, armano18}; this replacement is already planned, since the gold wire is the main limiting factor for MICROSCOPE's test of the WEP \cite{touboul17, touboul19}. Furthermore, patch effects will need to be either controlled or measurable, for instance by including a Kelvin probe in the instrument.
Finally, a possible Yukawa interaction at ranges $10^{-4}~{\rm m} \leqslant \lambda \leqslant 1~{\rm m}$ is expected to have a strengh $\alpha<10^{-4}$, corresponding to a stiffness seven orders of magnitude lower than the electrostatic stiffness. Since MICROSCOPE's capacitive control and measurement prevents us from using an electrostatic shield similar to that used by torsion pendulum experiments, a competitive experimental constrain will thus require a control of the instrument's theoretical model of one part in 10 millions. Whether this endeavour is possible remains an open question. Nevertheless, it could be circumvented by performing the measurement with several (more than two) voltages switching on and off different sets of electrodes to empirically determine the geometry dependence of the electrostatic stiffness.

In the meantime, we use the measurements presented in this paper to provide new constraints on the chameleon model in a companion paper \cite{pernotborras_inprep} based on Refs. \cite{pernotborras19, pernotborras20}.

\ack
We acknowledge useful discussions with Bruno Christophe and Bernard Foulon, and thank Vincent Lebat for comments on this article.
We acknowledge the financial support of CNES through the APR program (``GMscope+'' and ``Microscope 2" projects). MPB is supported by a CNES/ONERA PhD grant. This work uses technical details of the T-SAGE instrument, installed on the CNES-ESA-ONERA-CNRS-OCA-DLR-ZARM MICROSCOPE mission. This work is supported in part by the EU Horizon 2020 research and innovation programme under the Marie-Sklodowska grant No. 690575. This article is based upon work related to the COST Action CA15117 (CANTATA) supported by COST (European Cooperation in Science and Technology).

\appendix
\section{Test mass dynamics} \label{app_dynamics}

Equation (\ref{eq_measurement}) is an idealised version of the more realistic description of Ref. \cite{touboul20}. First, the sensor is not perfectly aligned with the satellite's frame, as described by the $[\theta]$ matrix
\begin{equation} \label{eq_GammaApp}
\vec\Gamma_{\rm cont | instr} = [\theta] \left( \vv{\Delta\Gamma}_{\Earth | \rm sat} + \vec\Gamma_{\rm kin | sat} + \frac{\vec{F}_{\rm ext | sat}}{M}+ \frac{\vec{F}_{\rm th | sat}}{M} \right) - \frac{\vec{F}_{\rm loc | instr}}{m} - \frac{\vec{F}_{\rm pa | instr}}{m},
\end{equation}
where the subscripts ``$|$instr" and ``$|$sat" mean that forces and accelerations are expressed in the instrument or satellite frame, respectively.

Moreover, the measured acceleration is given by the control acceleration (\ref{eq_GammaApp}) affected by the matrix $[A]$ containing the instrument's scale factors, by electrostatic parasitic forces (since the applied electrostatic forces are the sum of the measured and parasitic electrostatic forces $m \vec\Gamma_{\rm cont | instr} = \vec{F}_{\rm el} = \vec{F}_{\rm el, meas} + \vec{F}_{\rm elec, par}$), by the measurement bias $\vv{b}_0$ due to the read-out circuit and by noise $\vv{n}$:
\begin{equation}
\vv\Gamma_{\rm meas | instr} = \vv{b}_0 + [A] \left( \vv\Gamma_{\rm cont | instr} - \frac{\vv{F}_{\rm elec, par | instr}}{m}\right) + K_2 \left[\vv\Gamma_{\rm cont | sat} \right]^2 + \vv{n}.
\end{equation}

We can then wrap up and write the measured acceleration explicitly:
\begin{multline} \label{eq_kth_measacc}
\vv\Gamma_{\rm meas | instr} = \vv{B}_0 + [A] [\theta]\left( \vv{\Delta\Gamma}_{\Earth {\rm | sat}} + \vv\Gamma_{\rm kin | sat} + \frac{\vv{F}_{\rm ext | sat}}{M}+ \frac{\vv{F}_{\rm th | sat}}{M} \right) - [A] \frac{\vec{F}_{\rm loc | instr}}{m} \\
+ K_2 \left[\vv\Gamma_{\rm cont | sat} \right]^2 + \vv{n},
\end{multline}
where
\begin{equation}
\vv{B}_0 \equiv \vv{b}_0 - [A] \left(\frac{\vec{F}_{\rm pa | instr}}{m} + \frac{\vec{F}_{\rm elec,par | instr}}{m}\right)
\end{equation}
is the overall bias and $K_2$ is the quadratic factor accounting for non-linearities in the electronics. 

In this article, following the measurements of Ref. \cite{touboul19}, we assume that $[A] = [\theta] = {\rm Id}$ (Identity matrix), that the drag-free perfectly cancels the external forces and we ignore the quadratic factor (see Refs. \cite{touboul17, robert20, chhuncqg5}), so that our main measurement equation is
\begin{equation} 
\vv\Gamma_{\rm meas | instr} = \vv{B}_0 + \vv{\Delta\Gamma}_{\Earth {\rm | sat}} + \vv\Gamma_{\rm kin | sat} - \frac{\vec{F}_{\rm loc | instr}}{m} + \vv{n}.
\end{equation}

\section{Electric configurations} \label{app_hrmfrm}

MICROSCOPE can be used with two electric configurations: in the full-range mode (FRM), voltages are high enough to be able to acquire the test masses, while the high-resolution mode (HRM), with lower voltages, allows for an optimal control of the test masses. Tables \ref{tab_hrm} and \ref{tab_frm} summarise the corresponding voltages (which appear in Eq. \ref{eq_forcey}). See Ref. \cite{liorzou20} for details.

\begin{table}
\caption{High-resolution mode (HRM) electric configuration. All voltages are in V.}
\begin{center}
\begin{tabular}{cccccc}
\hline
& $V_d$ & $V_p$ & $V'_{px}$ & $V'_{py/z}$ & $V'_{p\phi}$ \\
\hline
IS1-SUEP & 5 & 5 & -5 & 2.5 & -10 \\
IS2-SUEP & 5 & 5 & 0 & 2.5 & -10 \\
IS1-SUREF & 5 & 5 & -5 & 2.5 & -10 \\
IS2-SUREF & 5 & 5 & -10 & 0 & -10 \\
\hline
\end{tabular}
\end{center}\label{tab_hrm}
\label{default}
\end{table}%

\begin{table}
\caption{Full-range mode (FRM) electric configuration. All voltages are in V.}
\begin{center}
\begin{tabular}{cccccc}
\hline
& $V_d$ & $V_p$ & $V'_{px}$ & $V'_{py/z}$ & $V'_{p\phi}$ \\
\hline
IS1-SUEP & 1 & 42 & 0 & 0 & 0\\
IS2-SUEP & 1 & 42 & 0 & 0 & 0\\
IS1-SUREF & 1 & 42 & 0 & 0 & 0 \\
IS2-SUREF & 1 & 42 & 0 & 0 & 0\\
\hline
\end{tabular}
\end{center}\label{tab_frm}
\label{default}
\end{table}%

\section{Discussion of Chhun et al. \cite{chhuncqg5} analysis} \label{ssect_chhun}

In Ref. \cite{chhuncqg5}, Chhun {\em et al.} compute the electrostatic stiffness in HRM, using the same measurement sessions as those used here, with a simple ratio of sines amplitudes. They neglect the local gravity stiffness and assume a negligible gold wire's stiffness $k_w\approx0$ and no velocity-dependent term ($\lambda_w=0$), and they fit the position and acceleration as
\begin{align} \label{eq_chhun1}
x(t) & = x_0 \sin(\omega t + \psi_x) \\
\Gamma(t) & = \Gamma_0 \sin(\omega t + \psi_\Gamma),
\end{align}
and infer $k_\epsilon = m \Gamma_0 / x_0 - m\omega^2$, with the implicit assumption that $\psi_x=\psi_\Gamma$. Table \ref{tab_chhun} sums up their results.

Two important points need to be highlighted. First, the stiffnesses estimated under the very restrictive assumptions of Ref. \cite{chhuncqg5} are close to (yet inconsistent with) the expected electrostatic stiffnesses (with an accuracy ranging from a few to a dozen percent, especially on the radial axes, see Table \ref{tab_results}). Second, the stiffness estimated on the radial axes are consistent with each other, thus showing a good degree of cylindrical symmetry; this symmetry is clearly expected for the electrostatic stiffness, but may seem accidental for the gold wires.
Unless coincidental, those facts hint towards a total stiffness indeed dominated by the electrostatic stiffness, with negligible other contributors (e.g. gold wires).

\begin{table}
\caption{Total stiffness (identified as the electrostatic stiffness) measured in Ref. \cite{chhuncqg5}. The expected values can be found in Table \ref{tab_results}.}
\begin{center}
\begin{tabular}{cccc}
\hline
& $k_{\epsilon,x}~[\times10^{-3}~{\rm N/m}]$ & $k_{\epsilon,y}~[\times10^{-2}~{\rm N/m}]$ & $k_{\epsilon,z}~[\times10^{-2}~{\rm N/m}]$ \\
\hline
IS1-SUEP & $1.396\pm0.003$ & $-1.494\pm0.001$ & $-1.478\pm0.002$ \\
IS2-SUEP & $0.639\pm0.002$ & $-6.424\pm0.001$ & $-6.310\pm0.001$ \\
IS1-SUREF & $0.837\pm0.003$ & $-1.515\pm0.000$ & $-1.514\pm0.000$ \\
IS2-SUREF &  $4.424\pm0.010$ & $-8.170\pm0.001$ & $-7.144\pm0.001$ \\
\hline
\end{tabular}
\end{center}\label{tab_chhun}
\label{default}
\end{table}%

It is instructive to consider Eq. (\ref{eq_recast}) in view of Ref. \cite{chhuncqg5} analysis. However, instead of assuming that the gold wire has no stiffness, we now assume that its quality factor $Q\gg 1$ (this is equivalent from the point of view of MICROSCOPE's test of the WEP, where only the ratio $k_w/Q$ enters the measurement). Thus assuming $\phi \rightarrow 0$, we re-write Eq. (\ref{eq_recast}) as (Taylor expanding the sine and cosine at first order in $\phi$)
\begin{equation} \label{eq_chhun2}
\Gamma(t) = \pm \sqrt{\kappa_0^2 + 2\kappa_0 \kappa_w + \kappa_w^2 (1+\phi^2)} \sin\left( \omega t + \psi - \arctan \frac{\kappa_w \phi}{\kappa_0 \kappa_w}\right),
\end{equation}
which tends to $\lim_{\phi \to 0} \Gamma(t) = \pm|\kappa_0 + \kappa_w| \sin(\omega t + \psi)$.
It is thus clear that using Eq. (\ref{eq_chhun1}), Ref. \cite{chhuncqg5} estimates the total stiffness. Nevertheless, a subtlety remains. Rigorously, although the phase in Eq. (\ref{eq_chhun2}) should be that of the excitation, $\psi=\psi_x$, which may (and does) differ from the phase of the acceleration $\psi_\Gamma$, Ref. \cite{chhuncqg5} assumes $\psi_x=\psi_\Gamma$ (which is consistent with the assumption that the gold wire has zero stiffness). Unfortunately, the experiment contradicts this assumption (at least on the radial axes). 

Relaxing the $\psi_x=\psi_\Gamma$ hypothesis of Eq. (\ref{eq_chhun1}), we find almost unchanged total stiffnesses (with percent-level modifications), but a small residual with a $\pi/2$ phase offset remains after removing the best fit from the acceleration. This remaining small signal may be the sign of a non-zero contribution of the gold wires. Indeed, Eq. (\ref{eq_recast}) shows that at first order, the amplitude of this residual signal is proportional to the $k_w/Q$ ratio of the gold wires (when assuming $\lambda_w=0$). Alas, this does not teach us anything about the absolute order of magnitude of either $k_w$ or $Q$.

\section{Radial electrostatic stiffness due to the Y electrodes} \label{app_estiffness}

\begin{figure}
\begin{center}
\includegraphics[width=.55\textwidth]{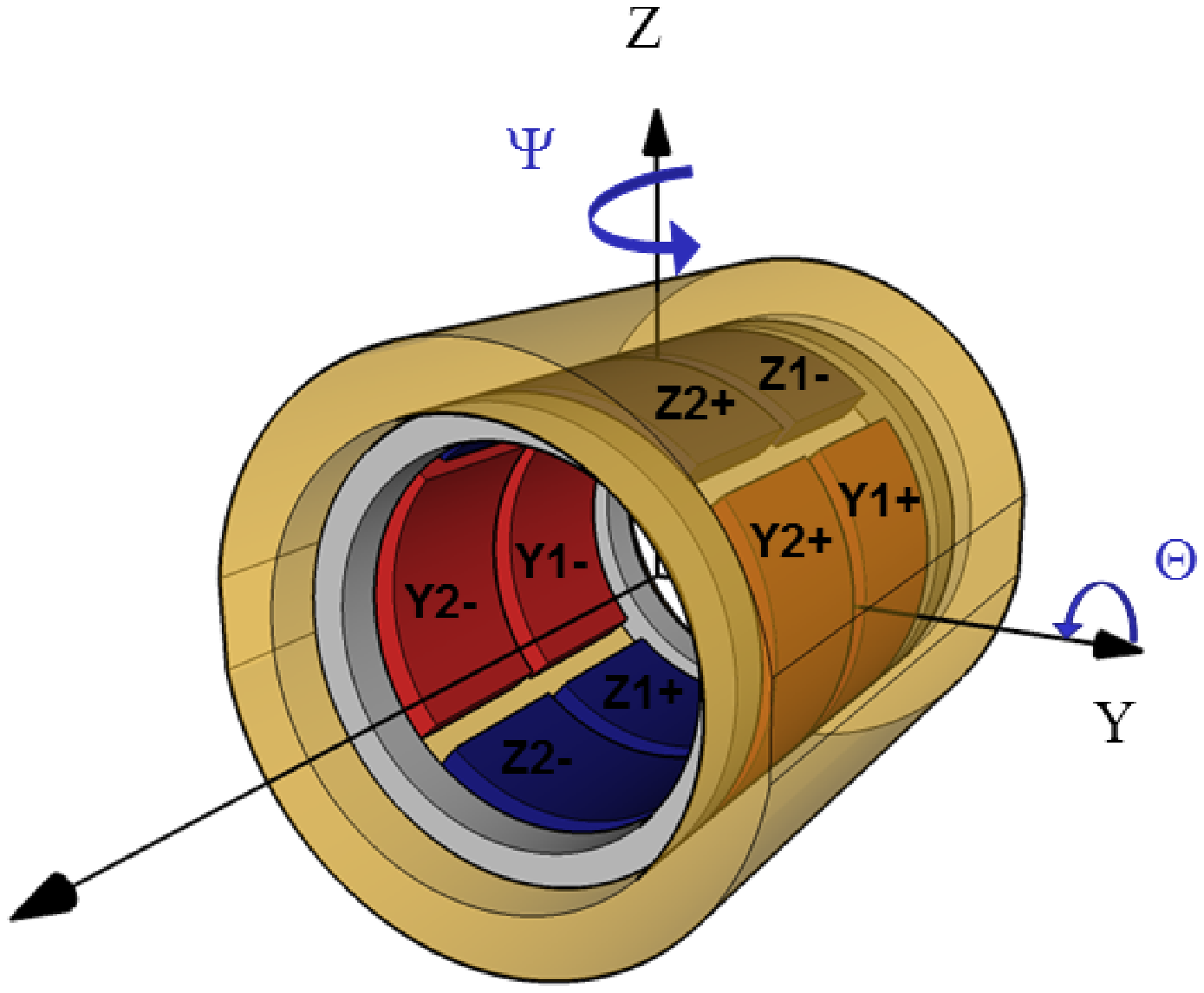}
\includegraphics[width=.55\textwidth]{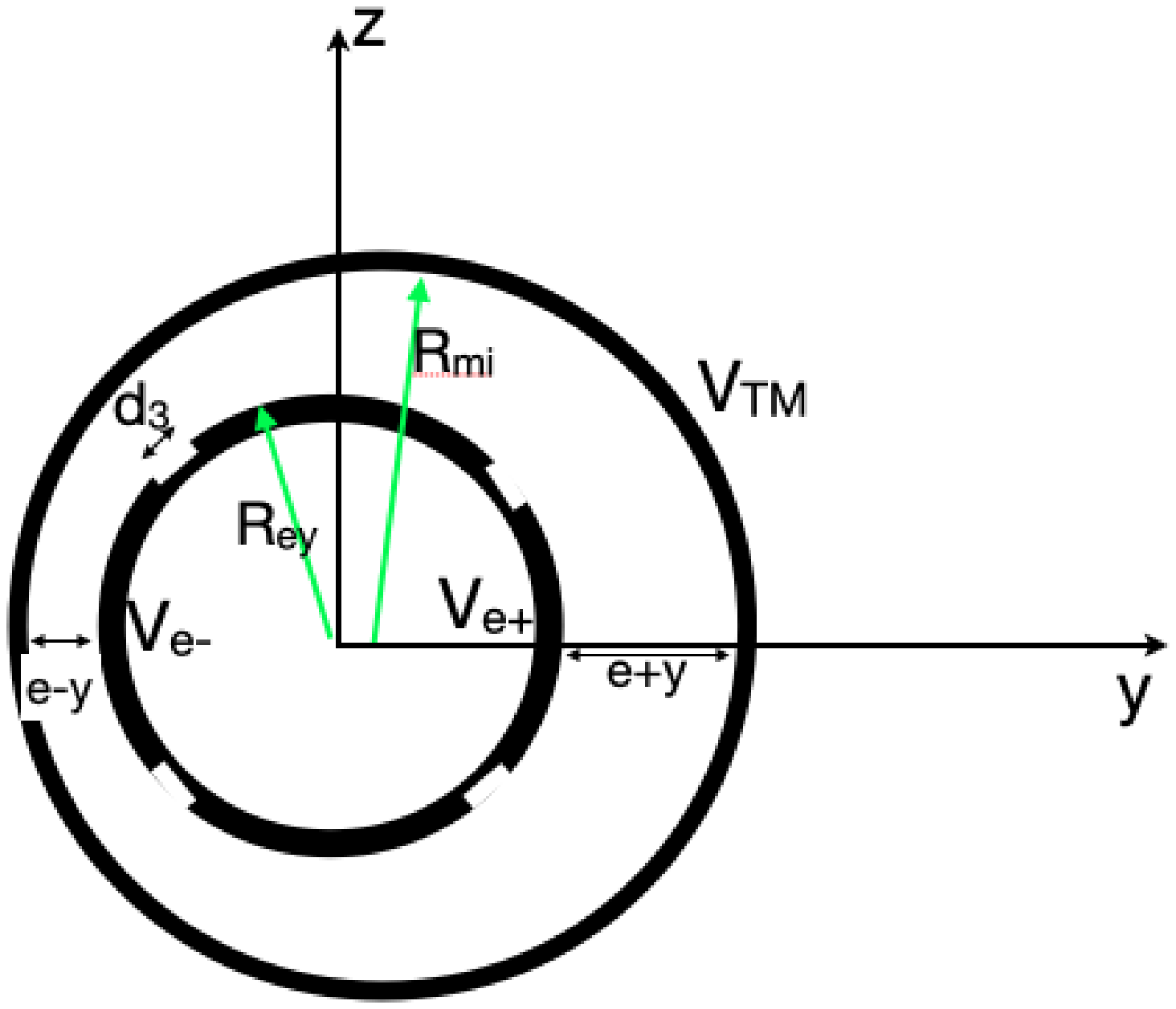}
\caption{Test mass' $Y$- and $Z$-axes control geometry. Upper panel: test mass (light brown) and inner electrode-bearing silica cylinder, with its two rings of pairs of electrodes to control the $Y$-axis (red) and the $Z-$axis (blue). The outer electrode-bearing silica cylinder controls the $X$-axis and is not shown here (see Ref. \cite{liorzou20}). Lower panel: Radial cut of a ring of $Y$ and $Z$ electrodes geometry, when the test mass is offset by $y$ along the $Y$-axis, with $e$ being the gap between the electrodes and the test mass in equilibrium. The inner cylinder carries the electrodes ($Y$ and $Z$ along the corresponding axes --$Y$ electrodes are shown at potential $V_{e+}$ and $V_{e-}$) of external radius $R_{ey}$; electrodes are separated by dips of width $d_3$. The test mass (of inner radius $R_{mi}$ and potential $V_{\rm TM}$) surrounds this inner cylinder, and can move around it.}
\label{fig_estiffness}
\end{center}
\end{figure}

In this appendix, we give a detailed computation of the electrostatic stiffness created by MICROSCOPE's $Y$ electrodes on a given test mass as the test mass moves along the $Y$-axis (but remains at $z=0$). Although this is textbook physics, this section allows us to clarify the model of the electrostatic stiffness. See Ref. \cite{hudson07} for the detailed general case. See Ref. \cite{liorzou20} for details about the geometry involved in this computation. In short, a given test mass is controled along its $Y$-axis by two pairs of diametrically-opposed $Y$ electrodes (at potential $V_{e+}$ and $V_{e-}$), completed by two pairs of $Z$-electrodes, as shown in Fig. \ref{fig_estiffness}.

\subsection{Electrostatic force between the plates of a capacitor}

At constant potential, the electrostatic force between conductors reads $F_{\rm elec} = \nabla U$, where $U$ is the electrostatic energy. For a capacitor,
\begin{equation}
U = \frac{1}{2} CV^2,
\end{equation} 
where $C$ is its capacitance and $V$ the potential difference between its plates. The electrostatic force created along the $y$-axis is then
\begin{equation} \label{eq_eforce}
F(y) = \frac{1}{2} \frac{\partial C}{\partial y} V^2.
\end{equation}

\subsection{Capacitance of one $Y$ electrode -- test mass pair}

Assuming electrodes are on an infinite cylinder (this assumption is reasonable since electrodes are far enough from the edges of the cylinder) and using the Gauss theorem, it is easy to show that the electric field of an electrode (of surface charge $\sigma$) at a distance $r$ from the axis of the cylinder is
\begin{equation}
E(r) = \frac{\sigma R_{ey}}{\epsilon_0 r}.
\end{equation}

The electric potential of the electrode is thus
\begin{equation}
V(r) = \frac{R_{ey} \sigma}{\epsilon_0} \ln r.
\end{equation}

Finally, the capacitance of the electrode-test mass pair 
\begin{equation} \label{eq_c1}
C = \frac{Q}{\Delta V} = \frac{1}{4} \left(2\pi - 4\frac{d_3}{R_{ey}}\right) \frac{L_y \epsilon_0}{\ln \frac{R_{mi}}{R_{ey}}},
\end{equation}
where the charge
\begin{equation} \label{eq_q}
Q = \sigma S = \frac{\sigma}{4} \left(2\pi - 4\frac{d_3}{R_{ey}}\right) R_{ey} L_y,
\end{equation}
where $S$ is the surface of an electrode (of length $L_y$).

Denoting $e \equiv R_{mi}-R_{ey}$ the gap between the cylinder and the test mass, in the limit $e \ll R_{ey} \sim R_{mi}$, Eq. (\ref{eq_c1}) reads
\begin{equation}
C = \frac{1}{4} \left(2\pi - 4 \frac{d_3}{R_{ey}}\right) L_y \epsilon_0 \frac{R_{mi} + R_{ey}}{2e}.
\end{equation}

\subsection{$Y$ electrodes electrostatic stiffness}

When moving the test mass by an amount $y$ along the $Y$-axis, the electrostatic force between the electrodes and the test mass is the sum of the forces between the test mass and the $V_{e+}$ and $V_{e-}$ electrodes, $F=F_+ + F_-$ (so far we consider only one pair of electrodes).

Those forces are, from Eq. (\ref{eq_eforce}),
\begin{equation}
F_+ = \frac{1}{2} \frac{\partial C_+}{\partial y} (V_{\rm TM} - V_{e+})^2,
\end{equation}
\begin{equation}
F_- = \frac{1}{2} \frac{\partial C_-}{\partial y} (V_{\rm TM} + V_{e-})^2,
\end{equation}
with
\begin{equation}
C_\pm = \frac{1}{4} \left(2\pi - 4 \frac{d_3}{R_{ey}}\right) L_y \epsilon_0 \frac{R_{mi} + R_{ey}}{2 (e \pm y)}.
\end{equation}

The total force is thus
\begin{equation}
F = k' \left[ -\frac{(V_{\rm TM} - V_{e+})^2}{(e+y)^2} + \frac{(V_{\rm TM} - V_{e-})^2}{(e-y)^2} \right],
\end{equation}
where
\begin{equation} \label{eq_kp}
k' \equiv \frac{1}{16} \left(2\pi - 4 \frac{d_3}{R_{ey}}\right) L_y \epsilon_0 (R_{mi} + R_{ey}).
\end{equation}

Assuming $y \ll e$, the force reads, at first order in $y/e$,
\begin{equation}
F = \frac{k'}{e^2} \left[ (V_{\rm TM} - V_{e-})^2 \left( 1+ 2\frac{y}{e}\right) - (V_{\rm TM} - V_{e+})^2 \left( 1-2\frac{y}{e}\right)\right].
\end{equation}

Keeping only the (stiffness) terms proportional to the displacement $y$ and expanding the square sums, we get
\begin{equation}
F = 2 \frac{k'}{e^3} \left[ -2 (V_{e+} + V_{e-}) V_{\rm TM} + V_{e-}^2 + V_{e+}^2 + 2V_{\rm TM}^2 \right] y,
\end{equation}
with \cite{liorzou20}
\begin{equation}
\left\{
\begin{array}{l}
V_{e-} = V'_p - v_y \\
V_{e+} = V'_p + v_y \\
V_{\rm TM} = V_p + \sqrt{2} V_d \sin \omega_d t,
\end{array}
\right.
\end{equation}
of which we take the mean value $\langle V_{\rm TM} \rangle = V_p$ and $\langle V_{\rm TM}^2 \rangle = V_p^2 + V_d^2$ (and omit the $\langle \dots \rangle$ symbol hereafter), such that the stiffness contribution to the force is
\begin{equation}
F = 4 \frac{k'}{e^3} \left[(V_p - V'_p)^2 + V_d^2\right] y.
\end{equation}

Considering now the two pairs of electrodes, and substituting Eq. (\ref{eq_kp}) to $k'$,
\begin{equation} \label{eq_Ftot}
F = \frac{1}{2} \left(2\pi - 4 \frac{d_3}{R_{ey}}\right) L_y \epsilon_0 \frac{R_{mi} + R_{ey}}{e^3} \left[(V_p - V'_p)^2 + V_d^2\right] y.
\end{equation}

Since $R_{mi} \approx R_{ey}$, using the expression for the surface of an electrode (Eq. \ref{eq_q}), we find the expression given in Eq. (\ref{eq_forcey}), with $\alpha_y=0$.

\section{Gravitational force between hollow cylinders} \label{appG}

Ref. \cite{lockerbie96} derives the longitudinal $F_z(r,z)$ and axial $F_r(r, z)$ forces between two hollow cylinders by a Yukawa gravitation. In this appendix, we use those results to complement them with the cases at hand in this paper. Note that contrary to the MICROSCOPE reference frame used in the main text, we use a more intuitive coordinate frame, where the $z$-axis is along the main axis of the cylinders, so that the natural cylindrical system $(r,\varphi,z)$ holds. This is the convention of Ref. \cite{lockerbie96}.

The gravitational force created along the $z$-axis on a unit mass at $(r,\theta,z)$ by a hollow cylinder of inner and outer radii $a$ and $b$, height $2\ell$ and density $\rho$ is \cite{lockerbie96}
\begin{equation} \label{eq_fz1}
F_z(r,z) = -2\pi G \alpha \rho \int_0^\infty \frac{J_0(kr) {\rm d}k}{\kappa} \left[b J_1(kb) - a J_1(ka)\right] \times
\left\{
\begin{array}{l}
h_2(z; k) \,\, {\rm if } \, -\ell \leqslant z \leqslant \ell \\
h_1(z; k) \,\, {\rm if } \, z > \ell \\ 
h_3(z; k) \,\, {\rm if } \, z < -\ell. \\
\end{array}
\right.
\end{equation}
where $\kappa$ is defined in Eq.~(\ref{eq_kappa_param}), with $\lambda$ the Yukawa interaction range, $J_i$ are Bessel functions of the first kind and the $h_i$ functions depend on the altitude of the unit mass and are defined below \footnote{Note that $h_1$ and $h_3$ are confused in Ref. \cite{lockerbie96}}. The Newtonian interaction is straightforward to recover by setting $\lambda \rightarrow \infty$ (and $\alpha=1$).

The corresponding radial force is given by
\begin{equation} \label{eq_fr1}
F_r(r,z) = -2\pi G \alpha \rho \int_0^\infty \frac{kJ_1(kr) {\rm d}k}{\kappa^2} \left[b J_1(kb) - a J_1(ka)\right] \times
\left\{
\begin{array}{l}
h_4(z; k) \,\, {\rm if } \, -\ell \leqslant z \leqslant \ell \\
h_1(z; k) \,\, {\rm if } \, z > \ell \\ 
-h_3(z; k) \,\, {\rm if } \, z < -\ell \\
\end{array}
\right.
\end{equation}

The $h_i$ functions are defined as
\begin{equation}
\begin{array}{l}
h_1(z;k) = \exp[-\kappa(z-\ell)] - \exp[-\kappa(\ell+z)] \\
h_2(z;k) = \exp[-\kappa(\ell-z)] - \exp[-\kappa(\ell+z)] \\
h_3(z;k) = \exp[\kappa(z-\ell)] - \exp[\kappa(\ell+z)] \\
h_4(z;k) = 2 - \exp[-\kappa(\ell-z)] - \exp[-\kappa(\ell+z)].
\end{array}
\end{equation}

\subsection{Forces on a full cylinder}

The forces exerted by the previous cylinder (called the ``source", centered on $(x,y,z)=(0,0,0)$) on another full cylinder (called the ``target", centered on $(x_s, 0, z_s)$) of radius $a'$, height $2L$ and density $\rho'$, is obtained by integrating Eqs. (\ref{eq_fz1}) and (\ref{eq_fr1}) on the volume of the target (at this point in the computation, we do not care whether the geometry is physically sound --i.e. cylinders may overlap; this will be done below):
\begin{equation} \label{eq_Fzg}
{\mathcal F_z}(x_s, z_s) = \rho' \iint {\rm d}x {\rm d}y \int_{z_{\rm min}}^{z_{\rm max}} {\rm d}z F_z(r,z),
\end{equation}
and similarly for ${\mathcal F_r}(x_s, z_s)$, where, for convenience, we express the volume in Cartesian coordinates (though we will quickly return to cylindrical coordinates below), with $r = \sqrt{x^2 + y^2}$. The $z$-integral is taken from the base $z_{\rm min}$ to the top $z_{\rm max}$ of the target cylinder, and the $(x,y)$-integral is taken over the disk section of the cylinder. We explicit them below.

\subsubsection{$z$-integral} \label{ssect_appzint}

$F_z(r,z)$ and $F_r(r,z)$ depend on $z$ only through the $h_i$ functions, so it is enough to compute $H_i(k) = \int_{z_{\rm min}}^{z_{\rm max}} h_i(z;k) {\rm d}z$. Several cases depending on the position of the target with respect to the source must be considered:

\begin{enumerate}

\item Target's $z$-extension fully contained in source's $z$-extension:
in this case, $z_s-L > -\ell$ and $z_s + L < \ell$, and only $h_2$ and $h_4$ are defined. Their integrals are straightforward to compute, with $z_{\rm min} = z_s - L$ and $z_{\rm max} = z_s + L$:

\begin{equation}
H_2(z_s, k) = \frac{4 {\rm e}^{-\kappa \ell}}{\kappa} \sinh(\kappa L) \sinh (\kappa z_s)
\end{equation}
and
\begin{equation}
H_4(z_s, k) = 4L - \frac{4 {\rm e}^{-\kappa \ell}}{\kappa} \sinh(\kappa L) \cosh (\kappa z_s).
\end{equation}

\item Target's $z$-extension fully covering source's $z$-extension ($z_s - L < -\ell$ and $z_s + L > \ell$): in this case, all $h_i$ are defined, and
\begin{equation}
H_1(z_s, k) = -\frac{2}{\kappa} \left({\rm e}^{-\kappa(z_s+L)} - {\rm e}^{-\kappa \ell} \right) \sinh(\kappa \ell),
\end{equation}
\begin{equation}
H_3(z_s, k) = -\frac{2}{\kappa} \left({\rm e}^{-\kappa \ell} - {\rm e}^{\kappa(z_s-L)} \right) \sinh(\kappa \ell),
\end{equation}
\begin{equation}
H_4(z_s, k) = 4\ell - 4 \frac{{\rm e}^{-\kappa \ell}}{\kappa} \sinh(\kappa \ell)
\end{equation}
and $H_2(z_s, k)=0$ by symmetry.

\item Target fully above source ($z_s - L > \ell$): in this case, only $h_1$ is defined and
\begin{equation}
H_1(z_s, k) = \frac{4 {\rm e}^{-\kappa z_s}}{\kappa} \sinh(\kappa \ell) \sinh(\kappa L).
\end{equation}

\item Target fully below source ($z_s + L < -\ell$): in this case, only $h_3$ is defined and
\begin{equation}
H_3(z_s, k) = -\frac{4 {\rm e}^{\kappa z_s}}{\kappa} \sinh(\kappa \ell) \sinh(\kappa L).
\end{equation}

\item Other cases correspond to the target's and the source's $z$-extension overlapping, with none completely covering the other. Since they are not of use in MICROSCOPE, we do not consider them here.

\end{enumerate}

\subsubsection{$(x,y)$-integral}

With no loss of generality, we can set the target cylinder on $(x_s, y_s) = (x_s, 0)$ in the $(x,y)$-plane ($y_s \neq 0$ cases are recovered by a simple rotation).
For an arbitrary function $f$,
\begin{equation}
\iint {\rm d}x {\rm d}y f(x,y) = \int_{\theta_-}^{\theta_+}{\rm d}\theta \int_{R_-(\theta)}^{R_+(\theta)} f(r,\theta) r {\rm d}r,
\end{equation}
where the integration boundaries depend on the geometry of the problem. Let us assume that the disk over which we take the integral is centered on $(x_s, y_s) = (x_s, 0)$ and has a radius $a$ (not to be confused with the radius of the source --which is of no use here).

\begin{enumerate}
\item{$|x_s|>a$}

This case is illustrated by the left panel of Fig. \ref{fig_intxy}. It is easy to show that the $\theta$ integral runs from $\theta_-=-\arcsin(a/|x_s|)$ to $\theta_+=\arcsin(a/|x_s|)$. For a given $\theta$ in that domain, the $r$-integration then runs from $R_-(\theta)$ to $R_+(\theta)$ which are solutions of the quadratic equation
\begin{equation}
R^2 - 2x_s R \cos\theta + x_s^2 - a^2 = 0,
\end{equation}
and are given by
\begin{equation}
R_\pm(\theta) = x_s \cos\theta \pm \sqrt{a^2 - x_s^2 \sin^2\theta}.
\end{equation}

\begin{figure}
\begin{center}
\includegraphics[width=.9\textwidth]{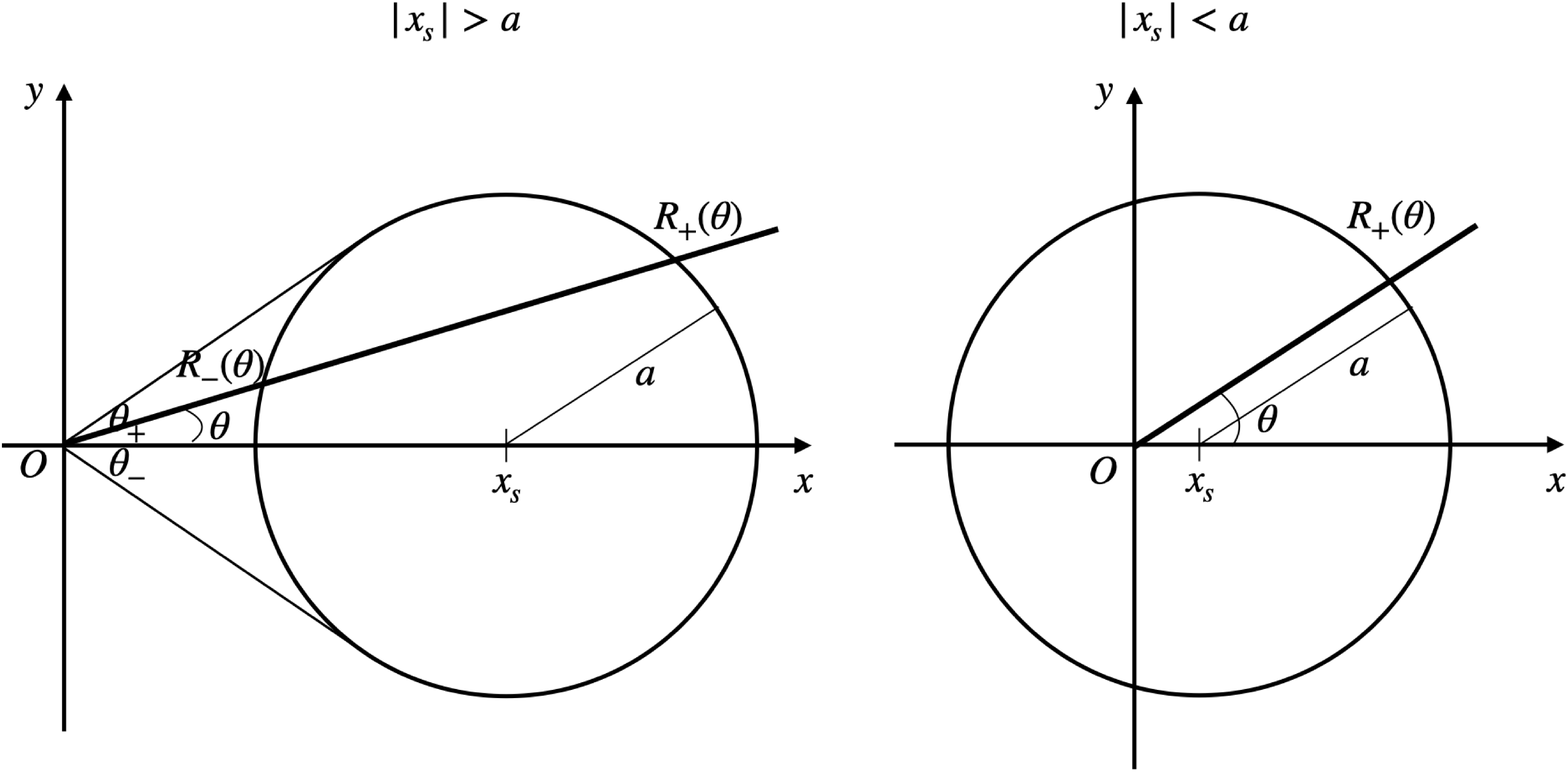}
\caption{$(x,y)$-integration geometry.}
\label{fig_intxy}
\end{center}
\end{figure}

\item{$|x_s| \leqslant a$}

In this case, shown in the right panel of Fig. \ref{fig_intxy}, the $\theta$ boundaries are trivially $\theta_-=0$ and $\theta_+=2\pi$. It is also trivial that for a given $\theta$, $R_-(\theta)=0$. Finally, it can be shown that the upper $r$-boundary is the same as that of the previous case, $R_+(\theta) = x_s \cos\theta + \sqrt{a^2 - x_s^2 \sin^2\theta}$.

\end{enumerate}

\subsubsection{Longitudinal and radial forces}

Noting that the $r$-dependence of the ${\mathcal F}_z(x_s,z_s)$ force appears only in the $J_0$ Bessel function, and using Eq. (\ref{eq_fz1}) we re-write Eq. (\ref{eq_Fzg}) as
\begin{equation}
{\mathcal F}_z(x_s, z_s) = -2\pi G \rho \rho' \alpha \int_0^\infty \frac{K_z(k)}{\kappa} \left[b J_1(kb) - a J_1(ka)\right] \left[H_1(z_s, k) + H_2(z_s, k) + H_3(z_s, k)\right] {\rm d}k,
\end{equation}
where we abusively sum the $H_i$ functions (setting them to 0 outside their definition range).

Since all cylinders of a given MICROSCOPE's sensor unit are co-axial, we consider only the $|x_s| \leqslant a$ case in this paper, so that the $(x,y)$-integration is 
\begin{equation} \label{eq_Kz}
K_z(k) = \int_0^{2\pi}{\rm d}\theta \int_0^{R_+(\theta)} J_0(kr) {\rm d}r,
\end{equation}
and we note that
\begin{equation}
I_z(k, \theta) \equiv \int_0^{R_+(\theta)} r J_0(kr) {\rm d}r = \frac{R_+(\theta) J_1[k R_+(\theta)]}{k}.
\end{equation}

Similarly, the radial force
\begin{equation}
{\mathcal F}_r(x_s, z_s) = -2\pi G \rho \rho' \alpha \int_0^\infty \frac{K_r(k)}{\kappa} \left[b J_1(kb) - a J_1(ka)\right] \left[H_1(z_s, k) + H_4(z_s, k) - H_3(z_s, k)\right] {\rm d}k,
\end{equation}
with 
\begin{equation} \label{eq_Kr}
K_r(k) = \int_0^{2\pi} \cos{\theta} {\rm d}\theta \int_0^{R_+(\theta)} kr J_1(kr) {\rm d}r,
\end{equation}
and
\begin{equation}
\begin{array}{lll}
I_r(k, \theta) & \equiv & \int_0^{R_+(\theta)} kr J_1(kr) {\rm d}r \\
 & = & \frac{\pi}{2} \left\{R_+(\theta) J_1[kR_+(\theta)] H_0[kR_+(\theta)] - R_+(\theta) J_0[kR_+(\theta)] H_1[kR_+(\theta)] \right\},
 \end{array}
\end{equation}
where $H_0$ and $H_1$ are Struve functions (not to be confused with the previous $H_i$ functions).

Without any further assumptions, we cannot integrate Eqs. (\ref{eq_Kz}) and (\ref{eq_Kr}) over $\theta$ analytically, and we end up with a 2D integral for the force between the two cylinders. We show below that in the limit of small displacements, we can integrate them analytically. Nevertheless, in the general case, the $\theta$ integrations are easily performed numerically.

\subsubsection{Small displacements limit: longitudinal force}

We assume that the target cylinder (of radius $a'$) moves about a ``reference" position ($\bar{x}, 0, \bar{z}$), with a small displacement $\delta$ along the $z$-axis. Assuming that $\bar{x}\ll a'$, at first order in $\bar{x}/a'$,
\begin{equation}
I_z(k, \theta) \approx a' \frac{J_1(ka')}{k} + a' J_0(ka') \cos\theta \bar{x},
\end{equation}
so that
\begin{equation}
K_z(k) \approx 2\pi \frac{a'J_1(ka')}{k}.
\end{equation}

Denoting $z_s = \bar{z} + \delta$ the altitude of the target's center, and expanding the $H_i$ functions in the limit of small $\delta$, taking care of their definition ranges (\ref{ssect_appzint}), we find that the longitudinal force created on the cylinder of radius $a'$ is, at third order:
\begin{enumerate}
\item if $\bar{z}\ll(\ell,L)$ and $\ell > L$ (target's $z$-extension fully covered by that of the source):
\begin{equation} \label{eq_fzTaylorAD}
{\mathcal F}_z(\bar{z}, \delta) \approx -16\pi^2 G \rho \rho' \alpha (K_1 \delta + K_3 \delta^3),
\end{equation}
where
\begin{equation}
K_1 = \int_0^\infty \frac{a'J_1(ka') [bJ_1(kb) - aJ_1(ka)]}{\kappa k} {\rm e}^{-\kappa \ell} \sinh(\kappa L) {\rm d}k
\end{equation}
and
\begin{equation}
K_3 = \int_0^\infty \frac{\kappa}{6} \frac{a'J_1(ka') [bJ_1(kb) - aJ_1(ka)]}{k} {\rm e}^{-\kappa \ell} \sinh(\kappa L) {\rm d}k.
\end{equation}

\item if $\bar{z}\ll(\ell,L)$ and $\ell < L$ (source's $z$-extension fully covered by that of the target): the force is formally identical to that of the previous case, with $\ell$ and $L$ switching their roles.

\item if $|\bar{z}| > \ell + L$ (cylinders above each other):
\begin{equation} \label{eq_fzTaylorB}
{\mathcal F}_z(\bar{z}, \delta) \approx 16\pi^2 G \rho \rho' \alpha (K_0 + K_1 \delta + K_2 \delta^2 + K_3 \delta^3),
\end{equation}
with
\begin{equation}
K_0 = - \frac{\bar{z}}{|\bar{z}|} \int_0^\infty \frac{a'J_1(ka') [bJ_1(kb) - aJ_1(ka)]}{\kappa^2 k} {\rm e}^{-\kappa |\bar{z}|} \sinh(\kappa \ell) \sinh(\kappa L) {\rm d}k,
\end{equation}
\begin{equation}
K_1 = \int_0^\infty \frac{a'J_1(ka') [bJ_1(kb) - aJ_1(ka)]}{\kappa k} {\rm e}^{-\kappa |\bar{z}|} \sinh(\kappa \ell) \sinh(\kappa L) {\rm d}k,
\end{equation}
\begin{equation}
K_2 = - \frac{\bar{z}}{|\bar{z}|} \int_0^\infty \frac{a'J_1(ka') [bJ_1(kb) - aJ_1(ka)]}{2 k} {\rm e}^{-\kappa |\bar{z}|} \sinh(\kappa \ell) \sinh(\kappa L) {\rm d}k,
\end{equation}
and
\begin{equation}
K_4 = \int_0^\infty \frac{\kappa}{6k} a'J_1(ka') [bJ_1(kb) - aJ_1(ka)] {\rm e}^{-\kappa |\bar{z}|} \sinh(\kappa \ell) \sinh(\kappa L) {\rm d}k.
\end{equation}

\end{enumerate}

\subsubsection{Small displacements limit: radial force}

We assume that the target cylinder (of radius $a'$) moves about a ``reference" position ($\bar{x}, 0, \bar{z}$), with a small displacement $\delta$ along the $X$-axis. Assuming that $\bar{x}\ll a'$, at third order in $\delta/a'$,
\begin{multline}
I_r(k, \theta) \approx \frac{\pi a'}{2} [J_1(ka') H_0(ka') - J_0(ka') H_1(ka')] + [ka' J_1(ka') \cos\theta] \delta \\
+ \frac{k}{2} [ka' J_0(ka') \cos^2\theta - J_1(ka') \sin^2\theta] \delta^2 \\
+ \frac{k^2}{6}[J_0(ka')\cos(3\theta) - ka'J_1(ka') \cos^3\theta] \delta^3,
\end{multline}
so that
\begin{equation}
K_r(k) \approx \pi k a' J_1(ka') \delta - \frac{\pi k^2}{8} ka' J_1(ka') \delta^3.
\end{equation}

The radial force created on the cylinder of radius $a'$ is thus, at third order
\begin{equation} \label{eq_frTaylor}
{\mathcal F}_r(\bar{z}, \delta) \approx -2\pi^2 G \rho \rho' \alpha (K_1 \delta + K_3 \delta^3),
\end{equation}
where, in the definition ranges (\ref{ssect_appzint}):
\begin{enumerate}
\item if $\bar{z}\ll(\ell,L)$ and $\ell > L$ (target's $z$-extension fully covered by that of the source):
\begin{equation}
K_1 = 4 \int_0^\infty \frac{k a'J_1(ka') [bJ_1(kb) - aJ_1(ka)]}{\kappa^2} \left[ L - \frac{{\rm e}^{-\kappa \ell}}{\kappa} \sinh(\kappa L) \cosh(\kappa \bar{z})\right] {\rm d}k
\end{equation}
and
\begin{equation}
K_3 = -\int_0^\infty \frac{k^3 a'J_1(ka') [bJ_1(kb) - aJ_1(ka)]}{\kappa^2} \left[ L - \frac{{\rm e}^{-\kappa \ell}}{\kappa} \sinh(\kappa L) \cosh(\kappa \bar{z})\right] {\rm d}k.
\end{equation}

\item if $\bar{z}\ll(\ell,L)$ and $\ell < L$ (source's $z$-extension fully covered by that of the target): the force is formally identical to that of the previous case, with $\ell$ and $L$ switching their roles.

\item if $|\bar{z}| > \ell + L$ (cylinders above each other):
\begin{equation}
K_1 = 4 \int_0^\infty \frac{k a'J_1(ka') [bJ_1(kb) - aJ_1(ka)]}{\kappa^2} \frac{{\rm e}^{-\kappa |\bar{z}|}}{\kappa} \sinh(\kappa \ell) \sinh(\kappa L) {\rm d}k
\end{equation}
\begin{equation}
K_3 = -\int_0^\infty \frac{k^3 a'J_1(ka') [bJ_1(kb) - aJ_1(ka)]}{\kappa^2} \frac{{\rm e}^{-\kappa |\bar{z}|}}{\kappa} \sinh(\kappa \ell) \sinh(\kappa L) {\rm d}k.
\end{equation}

\end{enumerate}

\subsection{Forces between hollow cylinders}

We finally come back to the problem at hand: the gravitational force between the two hollow cylinders defined at the beginning of this appendix. By virtue of the superposition principle, it is given by subtracting the force between the hollow source cylinder and two full target cylinders of radii $a$ and $b$. Thus, in the limit of small displacements, the longitudinal and radial forces are formally given by Eqs. (\ref{eq_fzTaylorAD}), (\ref{eq_fzTaylorB}) and (\ref{eq_frTaylor}), with the $K_i$ coefficients given below (they are obviously identical to those given in the main text in the MICROSCOPE coordinates system, where the $x$ and $z$-axes are inverted).

\subsubsection{Longitudinal force}

\begin{enumerate}
\item if $\bar{z}\ll(\ell,L)$ and $\ell > L$ (target's $z$-extension fully covered by that of the source):
\begin{equation} 
{\mathcal F}_z(\bar{z}, \delta) \approx -16\pi^2 G \rho \rho' \alpha (K_1 \delta + K_3 \delta^3),
\end{equation}
where
\begin{align}
K_1 &= \int_0^\infty \frac{[b'J_1(kb') - a'J_1(ka')] [bJ_1(kb) - aJ_1(ka)]}{\kappa k} {\rm e}^{-\kappa \ell} \sinh(\kappa L) {\rm d}k \\
K_3 &= \int_0^\infty \frac{\kappa}{6} \frac{[b'J_1(kb') - a'J_1(ka')] [bJ_1(kb) - aJ_1(ka)]}{k} {\rm e}^{-\kappa \ell} \sinh(\kappa L) {\rm d}k.
\end{align}

\item if $\bar{z}\ll(\ell,L)$ and $\ell < L$ (source's $z$-extension fully covered by that of the target): the force is formally identical to that of the previous case, with $\ell$ and $L$ switching their roles.

\item if $|\bar{z}| > \ell + L$ (cylinders above each other):
\begin{equation} 
{\mathcal F}_z(\bar{z}, \delta) \approx 16\pi^2 G \rho \rho' \alpha (K_0 + K_1 \delta + K_2 \delta^2 + K_3 \delta^3),
\end{equation}
with
\begin{align}
K_0 &= - \frac{\bar{z}}{|\bar{z}|} \int_0^\infty \frac{[b'J_1(kb') - a'J_1(ka')] [bJ_1(kb) - aJ_1(ka)]}{\kappa^2 k} {\rm e}^{-\kappa |\bar{z}|} \sinh(\kappa \ell) \sinh(\kappa L) {\rm d}k \\
K_1 &= \int_0^\infty \frac{[b'J_1(kb') - a'J_1(ka')] [bJ_1(kb) - aJ_1(ka)]}{\kappa k} {\rm e}^{-\kappa |\bar{z}|} \sinh(\kappa \ell) \sinh(\kappa L) {\rm d}k \\
K_2 &= - \frac{\bar{z}}{|\bar{z}|} \int_0^\infty \frac{[b'J_1(kb') - a'J_1(ka')] [bJ_1(kb) - aJ_1(ka)]}{2 k} {\rm e}^{-\kappa |\bar{z}|} \sinh(\kappa \ell) \sinh(\kappa L) {\rm d}k \\
K_4 &= \int_0^\infty \frac{\kappa}{6k} [b'J_1(kb') - a'J_1(ka')] [bJ_1(kb) - aJ_1(ka)] {\rm e}^{-\kappa |\bar{z}|} \sinh(\kappa \ell) \sinh(\kappa L) {\rm d}k,
\end{align}

\end{enumerate}

\subsubsection{Radial force}

\begin{equation} \label{eq_frTaylor2}
{\mathcal F}_r(\bar{z}, \delta) \approx -2\pi^2 G \rho \rho' \alpha (K_1 \delta + K_3 \delta^3),
\end{equation}

\begin{enumerate}
\item if $\bar{z}\ll(\ell,L)$ and $\ell > L$ (target's $z$-extension fully covered by that of the source):
\begin{align}
K_1 &= 4 \int_0^\infty \frac{k [b'J_1(kb') - a'J_1(ka')] [bJ_1(kb) - aJ_1(ka)]}{\kappa^2} \left[ L - \frac{{\rm e}^{-\kappa \ell}}{\kappa} \sinh(\kappa L) \cosh(\kappa \bar{z})\right] {\rm d}k \\
K_3 &= -\int_0^\infty \frac{k^3 [b'J_1(kb') - a'J_1(ka')] [bJ_1(kb) - aJ_1(ka)]}{\kappa^2} \left[ L - \frac{{\rm e}^{-\kappa \ell}}{\kappa} \sinh(\kappa L) \cosh(\kappa \bar{z})\right] {\rm d}k.
\end{align}

\item if $\bar{z}\ll(\ell,L)$ and $\ell < L$ (source's $z$-extension fully covered by that of the target): the force is formally identical to that of the previous case, with $\ell$ and $L$ switching their roles.

\item if $|\bar{z}| > \ell + L$ (cylinders above each other):
\begin{align}
K_1 &= 4 \int_0^\infty \frac{k [b'J_1(kb') - a'J_1(ka')] [bJ_1(kb) - aJ_1(ka)]}{\kappa^2} \frac{{\rm e}^{-\kappa |\bar{z}|}}{\kappa} \sinh(\kappa \ell) \sinh(\kappa L) {\rm d}k \\
K_3 &= -\int_0^\infty \frac{k^3 [b'J_1(kb') - a'J_1(ka')] [bJ_1(kb) - aJ_1(ka)]}{\kappa^2} \frac{{\rm e}^{-\kappa |\bar{z}|}}{\kappa} \sinh(\kappa \ell) \sinh(\kappa L) {\rm d}k.
\end{align}
\end{enumerate}

\subsection{MICROSCOPE gravitational stiffness}

\begin{figure}
\begin{center}
\includegraphics[width=.45\textwidth]{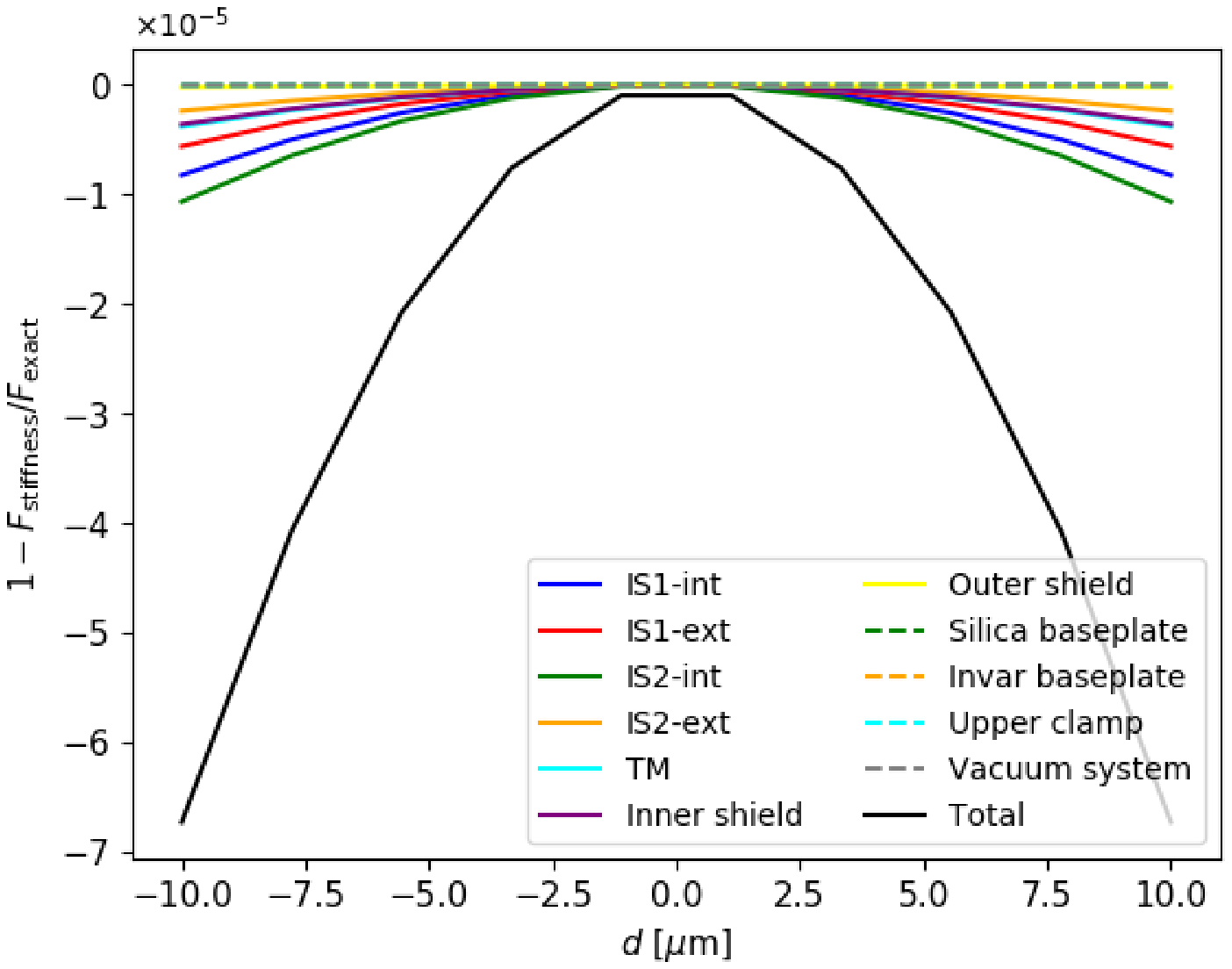}
\includegraphics[width=.45\textwidth]{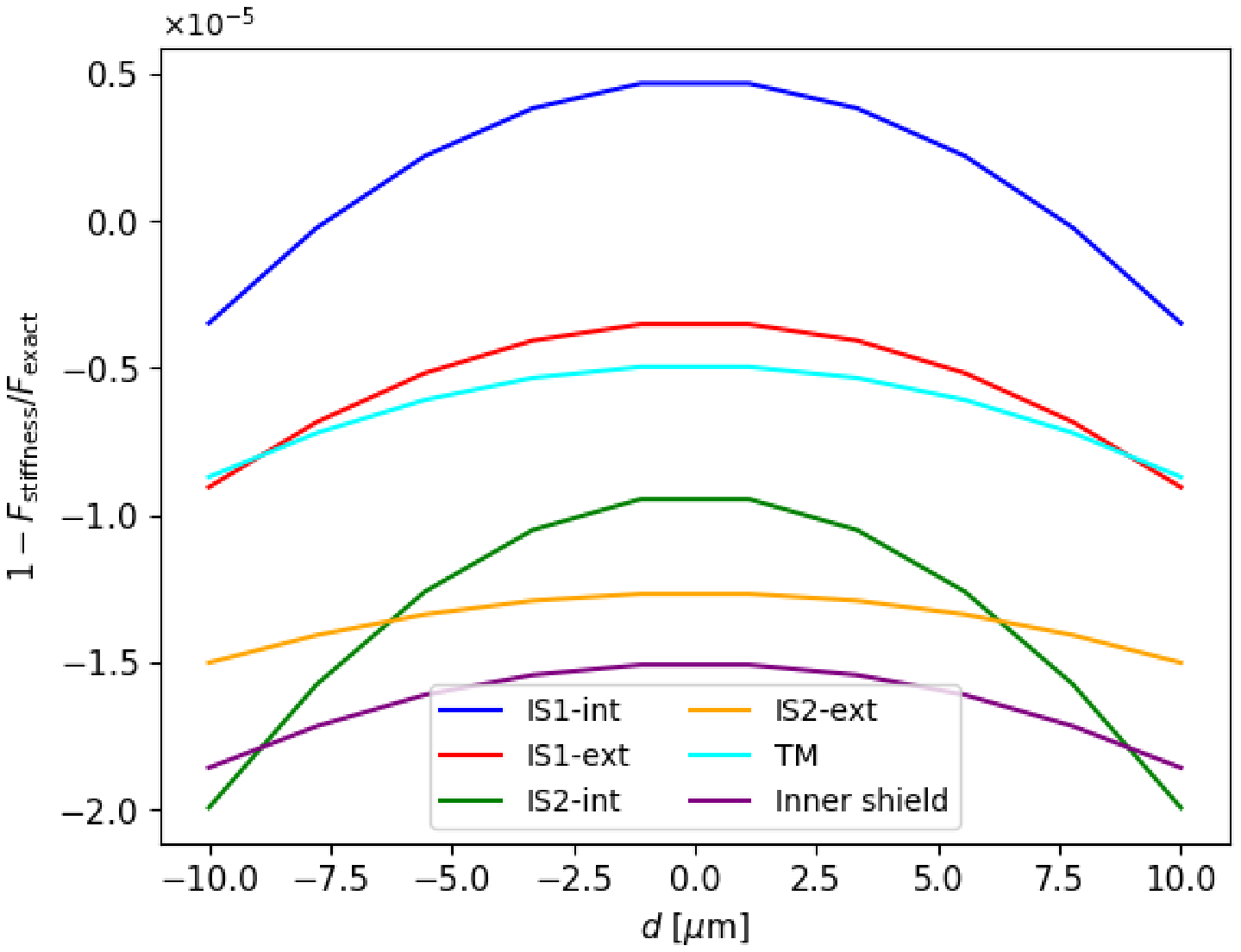}
\caption{Relative difference between the exact expression (\ref{eq_fr1}) and its first order Taylor expansion (first term of Eq. \ref{eq_frTaylor2}) for the radial force created by the parts of MICROSCOPE's SUEP on the inner test mass, as a function of the displacement of the test mass. Left: Newtonian force. Right: Yukawa force, for $(\alpha, 
\lambda) = (1, 0.01~{\rm m})$; only those cylinders which create a non-negligible Yukawa force allowing for a well-behaved $F_{\rm stiffness}/F_{\rm exact}$ ratio are shown.}
\label{fig_reldiff}
\end{center}
\end{figure}

Fig. \ref{fig_reldiff} shows the relative difference between the exact expression (\ref{eq_fr1}) and its first-order Taylor expansion (first term of Eq. \ref{eq_frTaylor2}) for the radial force created by the parts of MICROSCOPE's SUEP on the inner test mass, when the test mass moves within the range used to estimate the stiffness in flight. A first-order approximation provides a $10^{-5}$ accuracy on the gravitational forces, and can thus be safely used.

Finally, note that we compared the analytic results of this appendix with numerical simulations and with the (different) analytic expressions of Ref. \cite{hoyle04}, and found a good agreement between all methods.

\section*{References}
\bibliographystyle{iopart-num}
\bibliography{ystiff20}

\providecommand{\noopsort}[1]{}\providecommand{\singleletter}[1]{#1}%
\providecommand{\newblock}{}
\begin{thebibliography}{10}
\expandafter\ifx\csname url\endcsname\relax
  \def\url#1{{\tt #1}}\fi
\expandafter\ifx\csname urlprefix\endcsname\relax\def\urlprefix{URL }\fi
\providecommand{\eprint}[2][]{\url{#2}}

\bibitem{will14}
{Will} C~M 2014 {\em Living Reviews in Relativity\/} {\bf 17} 4
  (\textit{Preprint} \eprint{1403.7377})

\bibitem{bartelmann01}
{Bartelmann} M and {Schneider} P 2001 {\em Physics Reports\/} {\bf 340}
  291--472 (\textit{Preprint} \eprint{astro-ph/9912508})

\bibitem{hoekstra08}
{Hoekstra} H and {Jain} B 2008 {\em Annual Review of Nuclear and Particle
  Science\/} {\bf 58} 99--123 (\textit{Preprint} \eprint{0805.0139})

\bibitem{delva18}
{Delva} P, {Puchades} N, {Sch{\"o}nemann} E, {Dilssner} F, {Courde} C,
  {Bertone} S, {Gonzalez} F, {Hees} A, {Le Poncin-Lafitte} C, {Meynadier} F,
  {Prieto-Cerdeira} R, {Sohet} B, {Ventura-Traveset} J and {Wolf} P 2018 {\em
  Physical Review Letters\/} {\bf 121} 231101 (\textit{Preprint}
  \eprint{1812.03711})

\bibitem{herrmann18}
{Herrmann} S, {Finke} F, {L{\"u}lf} M, {Kichakova} O, {Puetzfeld} D,
  {Knickmann} D, {List} M, {Rievers} B, {Giorgi} G, {G{\"u}nther} C, {Dittus}
  H, {Prieto-Cerdeira} R, {Dilssner} F, {Gonzalez} F, {Sch{\"o}nemann} E,
  {Ventura-Traveset} J and {L{\"a}mmerzahl} C 2018 {\em Physical Review
  Letters\/} {\bf 121} 231102 (\textit{Preprint} \eprint{1812.09161})

\bibitem{abbott16}
Abbott B~P {\em et~al.\/} 2016 {\em Phys. Rev. Lett.\/} {\bf 116} 061102

\bibitem{zwicky33}
{Zwicky} F 1933 {\em Helvetica Physica Acta\/} {\bf 6} 110--127

\bibitem{rubin70}
{Rubin} V~C and {Ford} Jr W~K 1970 {\em Astrophysical Journal\/} {\bf 159} 379

\bibitem{riess98}
{Riess} A~G, {Filippenko} A~V, {Challis} P, {Clocchiatti} A, {Diercks} A,
  {Garnavich} P~M, {Gilliland} R~L, {Hogan} C~J, {Jha} S, {Kirshner} R~P,
  {Leibundgut} B, {Phillips} M~M, {Reiss} D, {Schmidt} B~P, {Schommer} R~A,
  {Smith} R~C, {Spyromilio} J, {Stubbs} C, {Suntzeff} N~B and {Tonry} J 1998
  {\em Astronomical Journal\/} {\bf 116} 1009--1038 (\textit{Preprint}
  \eprint{astro-ph/9805201})

\bibitem{perlmutter99}
{Perlmutter} S, {Aldering} G, {Goldhaber} G, {Knop} R~A, {Nugent} P, {Castro}
  P~G, {Deustua} S, {Fabbro} S, {Goobar} A, {Groom} D~E, {Hook} I~M, {Kim} A~G,
  {Kim} M~Y, {Lee} J~C, {Nunes} N~J, {Pain} R, {Pennypacker} C~R, {Quimby} R,
  {Lidman} C, {Ellis} R~S, {Irwin} M, {McMahon} R~G, {Ruiz-Lapuente} P,
  {Walton} N, {Schaefer} B, {Boyle} B~J, {Filippenko} A~V, {Matheson} T,
  {Fruchter} A~S, {Panagia} N, {Newberg} H~J~M, {Couch} W~J and {Project} T~S~C
  1999 {\em Astrophysical Journal\/} {\bf 517} 565--586 (\textit{Preprint}
  \eprint{astro-ph/9812133})

\bibitem{joyce15}
{Joyce} A, {Jain} B, {Khoury} J and {Trodden} M 2015 {\em Physics Reports\/}
  {\bf 568} 1--98 (\textit{Preprint} \eprint{1407.0059})

\bibitem{joyce16}
{Joyce} A, {Lombriser} L and {Schmidt} F 2016 {\em Annual Review of Nuclear and
  Particle Science\/} {\bf 66} 95--122 (\textit{Preprint} \eprint{1601.06133})

\bibitem{damour94}
{Damour} T and {Polyakov} A~M 1994 {\em Nucl. Phys. B\/} {\bf 423} 532--558
  (\textit{Preprint} \eprint{hep-th/9401069})

\bibitem{damour02}
{Damour} T, {Piazza} F and {Veneziano} G 2002 {\em Physical Review Letters\/}
  {\bf 89} 081601 (\textit{Preprint} \eprint{gr-qc/0204094})

\bibitem{damour92}
{Damour} T and {Esposito-Farese} G 1992 {\em Classical and Quantum Gravity\/}
  {\bf 9} 2093--2176

\bibitem{clifton12}
{Clifton} T, {Ferreira} P~G, {Padilla} A and {Skordis} C 2012 {\em Physics
  Reports\/} {\bf 513} 1--189 (\textit{Preprint} \eprint{1106.2476})

\bibitem{vainshtein72}
{Vainshtein} A~I 1972 {\em Physics Letters B\/} {\bf 39} 393--394

\bibitem{Damour:1992kf}
Damour T and Nordtvedt K 1993 {\em Phys. Rev. Lett.\/} {\bf 70} 2217--2219

\bibitem{khoury04a}
{Khoury} J and {Weltman} A 2004 {\em Phys. Rev. D\/} {\bf 69} 044026
  (\textit{Preprint} \eprint{astro-ph/0309411})

\bibitem{khoury04b}
{Khoury} J and {Weltman} A 2004 {\em Phys. Rev. Lett.\/} {\bf 93} 171104
  (\textit{Preprint} \eprint{astro-ph/0309300})

\bibitem{babichev09}
{Babichev} E, {Deffayet} C and {Ziour} R 2009 {\em Int. J. Mod. Phys. D\/} {\bf
  18} 2147--2154 (\textit{Preprint} \eprint{0905.2943})

\bibitem{hinterbichler10}
{Hinterbichler} K and {Khoury} J 2010 {\em Phys. Rev. Lett.\/} {\bf 104} 231301
  (\textit{Preprint} \eprint{1001.4525})

\bibitem{brax13}
{Brax} P, {Burrage} C and {Davis} A~C 2013 {\em JCAP\/} {\bf 1} 020
  (\textit{Preprint} \eprint{1209.1293})

\bibitem{burrage18}
{Burrage} C and {Sakstein} J 2018 {\em Living Reviews in Relativity\/} {\bf 21}
  1 (\textit{Preprint} \eprint{1709.09071})

\bibitem{damour12}
{Damour} T 2012 {\em Classical and Quantum Gravity\/} {\bf 29} 184001
  (\textit{Preprint} \eprint{1202.6311})

\bibitem{uzan03}
Uzan J~P 2003 {\em Rev. Mod. Phys.\/} {\bf 75} 403 (\textit{Preprint}
  \eprint{hep-ph/0205340})

\bibitem{uzan11}
Uzan J~P 2011 {\em Living Rev. Rel.\/} {\bf 14} 2 (\textit{Preprint}
  \eprint{1009.5514})

\bibitem{fischbach99}
{Fischbach} E and {Talmadge} C~L 1999 {\em {The Search for Non-Newtonian
  Gravity}\/}

\bibitem{adelberger03}
{Adelberger} E~G, {Heckel} B~R and {Nelson} A~E 2003 {\em Annual Review of
  Nuclear and Particle Science\/} {\bf 53} 77--121 (\textit{Preprint}
  \eprint{hep-ph/0307284})

\bibitem{adelberger09}
{Adelberger} E~G, {Gundlach} J~H, {Heckel} B~R, {Hoedl} S and {Schlamminger} S
  2009 {\em Progress in Particle and Nuclear Physics\/} {\bf 62} 102--134

\bibitem{tan20}
{Tan} W~H, {Du} A~B, {Dong} W~C, {Yang} S~Q, {Shao} C~G, {Guan} S~G, {Wang}
  Q~L, {Zhan} B~F, {Luo} P~S, {Tu} L~C and {Luo} J 2020 {\em Physical Review
  Letters\/} {\bf 124} 051301

\bibitem{lee20}
{Lee} J~G, {Adelberger} E~G, {Cook} T~S, {Fleischer} S~M and {Heckel} B~R 2020
  {\em Physical Review Letters\/} {\bf 124} 101101 (\textit{Preprint}
  \eprint{2002.11761})

\bibitem{touboul17}
{Touboul} P, {M{\'e}tris} G, {Rodrigues} M, {Andr{\'e}} Y, {Baghi} Q,
  {Berg{\'e}} J, {Boulanger} D, {Bremer} S, {Carle} P, {Chhun} R, {Christophe}
  B, {Cipolla} V, {Damour} T, {Danto} P, {Dittus} H, {Fayet} P, {Foulon} B,
  {Gageant} C, {Guidotti} P~Y, {Hagedorn} D, {Hardy} E, {Huynh} P~A,
  {Inchauspe} H, {Kayser} P, {Lala} S, {L{\"a}mmerzahl} C, {Lebat} V, {Leseur}
  P, {Liorzou} F, {List} M, {L{\"o}ffler} F, {Panet} I, {Pouilloux} B, {Prieur}
  P, {Rebray} A, {Reynaud} S, {Rievers} B, {Robert} A, {Selig} H, {Serron} L,
  {Sumner} T, {Tanguy} N and {Visser} P 2017 {\em Physical Review Letters\/}
  {\bf 119} 231101 (\textit{Preprint} \eprint{1712.01176})

\bibitem{touboul19}
{Touboul} P, {M{\'e}tris} G, {Rodrigues} M, {Andr{\'e}} Y, {Baghi} Q,
  {Berg{\'e}} J, {Boulanger} D, {Bremer} S, {Chhun} R, {Christophe} B,
  {Cipolla} V, {Damour} T, {Danto} P, {Dittus} H, {Fayet} P, {Foulon} B,
  {Guidotti} P~Y, {Hardy} E, {Huynh} P~A, {L{\"a}mmerzahl} C, {Lebat} V,
  {Liorzou} F, {List} M, {Panet} I, {Pires} S, {Pouilloux} B, {Prieur} P,
  {Reynaud} S, {Rievers} B, {Robert} A, {Selig} H, {Serron} L, {Sumner} T and
  {Visser} P 2019 {\em Classical and Quantum Gravity\/} {\bf 36} 225006
  (\textit{Preprint} \eprint{1909.10598})

\bibitem{berge18}
{Berg{\'e}} J, {Brax} P, {M{\'e}tris} G, {Pernot-Borr{\`a}s} M, {Touboul} P and
  {Uzan} J~P 2018 {\em Physical Review Letters\/} {\bf 120} 141101
  (\textit{Preprint} \eprint{1712.00483})

\bibitem{fayet18}
{Fayet} P 2018 {\em Phys. Rev. D\/} {\bf 97} 055039 (\textit{Preprint}
  \eprint{1712.00856})

\bibitem{fayet19}
Fayet P 2019 {\em Phys. Rev. D\/} {\bf 99}(5) 055043
  \urlprefix\url{https://link.aps.org/doi/10.1103/PhysRevD.99.055043}

\bibitem{mic20}
{Touboul} P in prep {\em Class. Quant. Grav.\/}

\bibitem{metris20}
{M\'etris} G in prep {\em Class. Quant. Grav.\/}

\bibitem{berge18cqg}
{Berg{\'e}} J, {Brax} P, {Pernot-Borr{\`a}s} M and {Uzan} J~P 2018 {\em
  Classical and Quantum Gravity\/} {\bf 35} 234001 (\textit{Preprint}
  \eprint{1808.00340})

\bibitem{kapner07}
{Kapner} D~J, {Cook} T~S, {Adelberger} E~G, {Gundlach} J~H, {Heckel} B~R,
  {Hoyle} C~D and {Swanson} H~E 2007 {\em Physical Review Letters\/} {\bf 98}
  021101 (\textit{Preprint} \eprint{hep-ph/0611184})

\bibitem{masuda09}
{Masuda} M and {Sasaki} M 2009 {\em Physical Review Letters\/} {\bf 102} 171101
  (\textit{Preprint} \eprint{0904.1834})

\bibitem{sushkov11}
{Sushkov} A~O, {Kim} W~J, {Dalvit} D~A~R and {Lamoreaux} S~K 2011 {\em Physical
  Review Letters\/} {\bf 107} 171101 (\textit{Preprint} \eprint{1108.2547})

\bibitem{klimchitskaya14}
{Klimchitskaya} G~L and {Mostepanenko} V~M 2014 {\em Gravitation and
  Cosmology\/} {\bf 20} 3--9 (\textit{Preprint} \eprint{1403.5778})

\bibitem{yang12}
{Yang} S~Q, {Zhan} B~F, {Wang} Q~L, {Shao} C~G, {Tu} L~C, {Tan} W~H and {Luo} J
  2012 {\em Physical Review Letters\/} {\bf 108} 081101

\bibitem{tan16}
{Tan} W~H, {Yang} S~Q, {Shao} C~G, {Li} J, {Du} A~B, {Zhan} B~F, {Wang} Q~L,
  {Luo} P~S, {Tu} L~C and {Luo} J 2016 {\em Physical Review Letters\/} {\bf
  116} 131101

\bibitem{touboul20}
{Touboul} P, {Rodrigues} M, {M{\'e}tris} G, {Chhun} R, {Robert} A, {Baghi} Q,
  {Hardy} E, {Berg{\'e}} J, {Boulanger} D, {Christophe} B, {Cipolla} V,
  {Foulon} B, {Guidotti} P~Y, {Huynh} P~A, {Lebat} V, {Liorzou} F, {Pouilloux}
  B, {Prieur} P and {Reynaud} S 2020 {\em arXiv e-prints\/} arXiv:2012.06472
  (\textit{Preprint} \eprint{2012.06472})

\bibitem{liorzou20}
{Liorzou} F, {Touboul} P, {Rodrigues} M, {M{\'e}tris} G, {Andr{\'e}} Y,
  {Berg{\'e}} J, {Boulanger} D, {Bremer} S, {Chhun} R, {Christophe} B, {Danto}
  P, {Foulon} B, {Hagedorn} D, {Hardy} E, {Huynh} P~A, {L{\"a}mmerzahl} C,
  {Lebat} V, {List} M, {L{\"o}ffler} F, {Rievers} B, {Robert} A and {Selig} H
  2020 {\em arXiv e-prints\/} arXiv:2012.11232 (\textit{Preprint}
  \eprint{2012.11232})

\bibitem{robert20}
{Robert} A, {Cipolla} V, {Prieur} P, {Touboul} P, {M{\'e}tris} G, {Rodrigues}
  M, {Andr{\'e}} Y, {Berg{\'e}} J, {Boulanger} D, {Chhun} R, {Christophe} B,
  {Guidotti} P~Y, {Hardy} E, {Lebat} V, {Lienart} T, {Liorzou} F and
  {Pouilloux} B 2020 {\em arXiv e-prints\/} arXiv:2012.06479 (\textit{Preprint}
  \eprint{2012.06479})

\bibitem{damour10a}
{Damour} T and {Donoghue} J~F 2010 {\em Physical Review D\/} {\bf 82} 084033
  (\textit{Preprint} \eprint{1007.2792})

\bibitem{damour10b}
{Damour} T and {Donoghue} J~F 2010 {\em Class. Quant. Grav.\/} {\bf 27} 202001
  (\textit{Preprint} \eprint{1007.2790})

\bibitem{chhuncqg5}
{Chhun} R in prep {\em Class. Quant. Grav.\/}

\bibitem{berge_cqg7}
{Berg{\'e}} J, {Baghi} Q, {Hardy} E, {M{\'e}tris} G, {Robert} A, {Rodrigues} M,
  {Touboul} P, {Chhun} R, {Guidotti} P~Y, {Pires} S, {Reynaud} S, {Serron} L
  and {Travert} J~M 2020 {\em arXiv e-prints\/} arXiv:2012.06484
  (\textit{Preprint} \eprint{2012.06484})

\bibitem{saulson90}
{Saulson} P~R 1990 {\em Physical Review D\/} {\bf 42} 2437--2445

\bibitem{willemenot00}
{Willemenot} E and {Touboul} P 2000 {\em Review of Scientific Instruments\/}
  {\bf 71} 302--309

\bibitem{nofrarias07}
{Nofrarias i Serra} M 2007 {\em Thermal Diagnostics in the LISA Technology
  Package Experiment\/} Ph.D. thesis Departament de F\'isica Fonamental.
  Universitat de Barcelona

\bibitem{carbone07}
{Carbone} L, {Cavalleri} A, {Ciani} G, {Dolesi} R, {Hueller} M, {Tombolato} D,
  {Vitale} S and {Weber} W~J 2007 {\em Physical Review D\/} {\bf 76} 102003
  (\textit{Preprint} \eprint{0706.4402})

\bibitem{hardycqg6}
{Hardy} E~~a TBP {\em Class. Quant. Grav.\/}

\bibitem{speake96}
{Speake} C~C 1996 {\em Classical and Quantum Gravity\/} {\bf 13} A291--A297

\bibitem{hudson07}
Hudson D 2007 {\em Investigation exp{\'e}rimentale et th{\'e}orique du
  prototype du capteur inertiel pour la v{\'e}rification du principe
  d'{\'e}quivalence dans la mission spatiale MICROSCOPE\/} Ph.D. thesis
  Universit{\'e} Pierre et Marie Curie

\bibitem{touboul09}
{Touboul} P 2009 {\em Space Science Reviews\/} {\bf 148} 455--474

\bibitem{hoskins85}
{Hoskins} J~K, {Newman} R~D, {Spero} R and {Schultz} J 1985 {\em Physical
  Review D\/} {\bf 32} 3084--3095

\bibitem{sumner09}
{Sumner} T~J, {Shaul} D~N~A, {Schulte} M~O, {Waschke} S, {Hollington} D and
  {Ara{\'u}jo} H 2009 {\em Classical and Quantum Gravity\/} {\bf 26} 094006

\bibitem{armano18}
{Armano} M, {Audley} H, {Baird} J, {Binetruy} P, {Born} M, {Bortoluzzi} D,
  {Castelli} E, {Cavalleri} A, {Cesarini} A, {Cruise} A~M, {Danzmann} K, {de
  Deus Silva} M, {Diepholz} I, {Dixon} G, {Dolesi} R, {Ferraioli} L, {Ferroni}
  V, {Fitzsimons} E~D, {Freschi} M, {Gesa} L, {Gibert} F, {Giardini} D,
  {Giusteri} R, {Grimani} C, {Grzymisch} J, {Harrison} I, {Heinzel} G,
  {Hewitson} M, {Hollington} D, {Hoyland} D, {Hueller} M, {Inchausp{\'e}} H,
  {Jennrich} O, {Jetzer} P, {Karnesis} N, {Kaune} B, {Korsakova} N, {Killow}
  C~J, {Lobo} J~A, {Lloro} I, {Liu} L, {L{\'o}pez-Zaragoza} J~P,
  {Maarschalkerweerd} R, {Mance} D, {Meshksar} N, {Mart{\'\i}n} V,
  {Martin-Polo} L, {Martino} J, {Martin-Porqueras} F, {Mateos} I, {McNamara}
  P~W, {Mendes} J, {Mendes} L, {Nofrarias} M, {Paczkowski} S, {Perreur-Lloyd}
  M, {Petiteau} A, {Pivato} P, {Plagnol} E, {Ramos-Castro} J, {Reiche} J,
  {Robertson} D~I, {Rivas} F, {Russano} G, {Slutsky} J, {Sopuerta} C~F,
  {Sumner} T, {Texier} D, {Thorpe} J~I, {Vetrugno} D, {Vitale} S, {Wanner} G,
  {Ward} H, {Wass} P~J, {Weber} W~J, {Wissel} L, {Wittchen} A and {Zweifel} P
  2018 {\em Physical Review Letters\/} {\bf 120} 061101

\bibitem{pernotborras_inprep}
{Pernot-Borr{\`a}s} M, {Berg{\'e}} J, {Brax} P and {Uzan} J~P in prep {\em
  Physical Review D\/}

\bibitem{pernotborras19}
{Pernot-Borr{\`a}s} M, {Berg{\'e}} J, {Brax} P and {Uzan} J~P 2019 {\em
  Physical Review D\/} {\bf 100} 084006 (\textit{Preprint} \eprint{1907.10546})

\bibitem{pernotborras20}
{Pernot-Borr{\`a}s} M, {Berg{\'e}} J, {Brax} P and {Uzan} J~P 2020 {\em
  Physical Review D\/} {\bf 101} 124056 (\textit{Preprint} \eprint{2004.08403})

\bibitem{lockerbie96}
{Lockerbie} N~A, {Veryaskin} A~V and {Xu} X 1996 {\em Journal of Physics A
  Mathematical General\/} {\bf 29} 4649--4663

\bibitem{hoyle04}
{Hoyle} C~D, {Kapner} D~J, {Heckel} B~R, {Adelberger} E~G, {Gundlach} J~H,
  {Schmidt} U and {Swanson} H~E 2004 {\em Physical Review D\/} {\bf 70} 042004
  (\textit{Preprint} \eprint{hep-ph/0405262})

\end{thebibliography}

\end{document}